\documentclass[structabstract]{aa}
\usepackage{graphicx}
\usepackage{natbib}
\usepackage{txfonts}
\begin{document}
\title{Searching for faint companions with VLTI/PIONIER}
\subtitle{I. Method and first results \thanks{Based on observations obtained at the European Southern Observatory (ESO) Very Large Telescope Interferometer (VLTI), Paranal, Chile.}}

\author{O. Absil\inst{1}\fnmsep\thanks{Postdoctoral Researcher F.R.S.-FNRS (Belgium), email: \texttt{absil@astro.ulg.ac.be}.}, J.-B. Le Bouquin\inst{2}, J.-P. Berger\inst{3}, A.-M. Lagrange\inst{2}, G. Chauvin\inst{2,4}, B. Lazareff\inst{2}, G. Zins\inst{2}, P. Haguenauer\inst{3}, L. Jocou\inst{2}, P. Kern\inst{2}, R. Millan-Gabet\inst{5}, S. Rochat\inst{2}, W. Traub\inst{6}}

\authorrunning{Absil et al.}

\institute{D\'epartement d'Astrophysique, G\'eophysique et Oc\'eanographie, Universit\'e de Li\`ege, 17 All\'ee du Six Ao\^ut, 4000 Li\`ege, Belgium
\and
Institut de Plan\'etologie et d'Astrophysique de Grenoble (IPAG) UMR 5274, UJF-Grenoble 1 / CNRS-INSU, Grenoble, France
\and
European Southern Observatory, Casilla 19001, Santiago 19, Chile
\and
Max-Planck-Institut f\"ur Astronomie, K\"onigstuhl 17, 69117 Heidelberg, Germany
\and
NASA Exoplanet Science Institute (NExScI), California Institute of Technology, Pasadena, CA 91125, USA
\and
Jet Propulsion Laboratory, California Institute of Technology, Pasadena, CA 91109, USA
}

\date{Received 18 July 2011; accepted 22 September 2011}

\abstract
{A new four-telescope interferometric instrument called PIONIER has recently been installed at VLTI. It provides improved imaging capabilities together with high precision.}
{We search for low-mass companions around a few bright stars using different strategies, and determine the dynamic range currently reachable with PIONIER.}
{Our method is based on the closure phase, which is the most robust interferometric quantity when searching for faint companions. We computed the $\chi^2$ goodness of fit for a series of binary star models at different positions and with various flux ratios. The resulting $\chi^2$ cube was used to identify the best-fit binary model and evaluate its significance, or to determine upper limits on the companion flux in case of non detections.}
{No companion is found around \object{Fomalhaut}, \object{tau Cet} and \object{Regulus}. The median upper limits at $3\sigma$ on the companion flux ratio are respectively of $2.3\times10^{-3}$ (in 4 hours), $3.5\times10^{-3}$ (in 3 hours) and $5.4\times10^{-3}$ (in 1.5 hour) on the search region extending from 5 to 100\,mas. Our observations confirm that the previously detected near-infrared excess emissions around Fomalhaut and tau Cet are not related to a low-mass companion, and instead come from an extended source such as an exozodiacal disk. In the case of \object{del Aqr}, in 30~min of observation, we obtain the first direct detection of a previously known companion, at an angular distance of about 40\,mas and with a flux ratio of $2.05\times10^{-2} \pm 0.16\times10^{-2}$. Due to the limited $u,v$ plane coverage, its position can, however, not be unambiguously determined.}
{After only a few months of operation, PIONIER has already achieved one of the best dynamic ranges world-wide for multi-aperture interferometers. A dynamic range up to about 1:500 is demonstrated, but significant improvements are still required to reach the ultimate goal of directly detecting hot giant extrasolar planets.}

\keywords{techniques: interferometric -- binaries: close -- stars: low mass -- brown dwarfs -- planetary systems}

   \maketitle
%

\section{Introduction} \label{sec:intro}

Searching for faint companions around bright nearby stars is one of the most challenging goals of today's astronomy \citep{Oppenheimer09,Absil10b}. This quest for high-dynamic range observations has been spurred for more than a decade by the discovery of more than 500 extrasolar planets in our galactic neighbourhood. A proper characterisation of the physical and atmospheric parameters of the detected planets requires the photon emitted (or reflected) by the planets to be disentangled from those emitted by their host star. While a temporal separation of the photons is possible in the particular case of transiting systems, most systems require high-dynamic range imaging with a high angular resolution in order to be characterised.

Adaptive optics on 10-m class telescopes, possibly aided by the use of a coronagraph and/or differential imaging techniques, is currently the most widely used technique for directly imaging extrasolar planets \citep[e.g.,][]{Chauvin04,Kalas08,Marois08,Lagrange09a}. This technique provides very high-contrast images at relatively large angular distances from the central star, up to about 1:$10^6$ beyond $1\arcsec$, and can detect companions with flux ratios around 1:1000 down to about 100\,mas separations \citep[e.g.][]{Boccaletti09}. The upcoming generation of high-contrast imagers using extreme adaptive optics on 10-m class telescope promises to further improve these performances. Combined with state-of-the-art coronagraphic devices such as vortex coronagraphs, the detection of high-contrast companions at the diffraction limit ($\lambda/D$, i.e., about 50\,mas at K band on a 10-m telescope) will be enabled \citep[e.g.,][]{Mawet11}.

Pushing the discovery space within the diffraction limit of a single aperture requires interferometric techniques. On a single pupil, aperture-masking techniques have recently allowed dynamic ranges up to 1:1000 to be reached down to the diffraction limit, and permitted decent contrasts ($\sim$\,1:100) to be reached down to $\lambda/3D$ \citep{Hinkley11,Lacour11}. Another example is the Palomar Fiber Nuller, which also works on a single telescope, and has reached a 1:500 dynamic range around the bright star Vega in a search region extending significantly within the diffraction limit of the Palomar Hale 5-m telescope \citep{Mennesson11}. To extend the search region even closer to the central star, interferometry with multiple apertures separated by more than 10\,m is mandatory. Interferometry on multiple apertures has, however, not yet reached the same dynamic range as adaptive optics or even aperture-masking observations, at least when the target star is mostly unresolved \citep{Absil10a}. The dynamic range can be significantly increased by observing resolved targets \citep[e.g.,][]{Duvert10,Zhao11}, but such targets are scarce in the context of main sequence stars observed on 100-m class baselines.

Here, we present the first results of a new 4-telescope interferometric instrument called PIONIER (Precision Integrated-Optics Near-infrared Imaging ExpeRiment), which has recently been installed and commissioned at the Very Large Telescope Interferometer \citep{LeBouquin11}. PIONIER aims at providing enhanced imaging capabilities at the VLTI, together with an improved accuracy and stability of interferometric observables (visibility and closure phase). One of its main scientific goals is to directly detect faint companions around bright stars (up to a magnitude $H\sim7$). The purpose of this paper is to discuss the dynamic range of this new instrument based on the observations obtained during the commissioning phase and the first scientific observing runs.


\section{Observations and data reduction}

\begin{table}[t]
\caption{Observing log for Fomalhaut, tau Cet, del Aqr, and Regulus.}
\centering
\begin{tabular}{c c c c c}
\hline\hline
Night & MJD & Files & Star & Baseline \\
\hline
2010 Nov 30 & 55530.999 & 3 & HD\,209688 & E0-G0-H0-I1 \\
... & 55531.007 & 10 & Fomalhaut & ... \\
... & 55531.023 & 5 & HD\,209688 & ...\\
... & 55531.035 & 10 & Fomalhaut & ...\\
... & 55531.048 & 5 & HD\,219784 & ... \\
... & 55531.055 & 10 & Fomalhaut & ...\\
... & 55531.067 & 10 & HD\,219784 & ...\\
... & 55531.098 & 10 & Fomalhaut & ...\\
... & 55531.114 & 10 & HD\,219784 & ...\\
2010 Dec 6 & 55537.047 & 5 & HD\,219784 & D0-G1-H0-I1 \\
... & 55537.058 & 5 & Fomalhaut & ... \\
... & 55537.067 & 5 & HD\,219784 & ... \\
... & 55537.076 & 5 & Fomalhaut & ... \\
... & 55537.084 & 5 & HD\,219784 & ... \\
... & 55537.090 & 5 & Fomalhaut & ... \\
... & 55537.096 & 5 & HD\,219784 & ... \\
\hline
2010 Dec 3 & 55534.053 & 5 & tau Cet & E0-G0-H0-I1 \\
... & 55534.060 & 5 & HD\,8705 & ... \\
... & 55534.066 & 5 & tau Cet & ... \\
... & 55534.075 & 5 & HD\,14376 & ... \\
... & 55534.081 & 5 & tau Cet & ... \\
... & 55534.092 & 5 & HD\,15694 & ... \\
... & 55534.106 & 5 & HD\,16526 & ... \\
... & 55534.113 & 5 & tau Cet & ... \\
... & 55534.121 & 5 & HD\,15694 & ... \\
2010 Dec 19 & 55550.031 & 5 & HD\,8705 & A0-G1-K0-I1 \\
... & 55550.041 & 5 & tau Cet & ... \\
... & 55550.051 & 5 & HD\,14376 & ... \\
... & 55550.073 & 5 & tau Cet & ... \\
... & 55550.083 & 5 & HD\,8705 & ... \\
\hline
2010 Oct 28 & 55498.089 & 3 & HD\,221745 & D0-E0-H0-I1 \\
... & 55498.096 & 3 & del Aqr & ... \\
... & 55498.105 & 3 & gam~Tuc & ... \\
\hline
2010 Nov 28 & 55529.319 & 4 & HD\,45018 & E0-G0-H0-I1 \\
... & 55529.335 & 5 & Regulus & ... \\
... & 55529.343 & 3 & HD\,83425 & ... \\
... & 55529.351 & 3 & Regulus & ... \\
... & 55529.359 & 3 & HD\,82381 & ... \\
... & 55529.369 & 3 & Regulus & ... \\
\hline
\end{tabular}
\label{tab:obs}
\end{table}

The observations presented in this paper were obtained during the commissioning phase and the first scientific observing runs of PIONIER at VLTI, which took place in Fall 2010. A large number of stars has been observed during several nights from October 25 to December 25, using only the 1.8-m Auxiliary Telescopes of the VLTI. Within this sample, we selected four stars that are representative of two possible observing strategy: either a long integration for a deep search for companions (cases of \object{Fomalhaut} and \object{tau Cet}) or a short (``snapshot'') integration for a shallow survey-type search (cases of \object{del Aqr} and \object{Regulus}). The observing log related to these four stars is displayed in Table~\ref{tab:obs}. Observations of the science targets were interleaved with observations of reference stars to calibrate the instrumental contribution in the observed quantities. Calibrators were usually chosen close to the science targets, both in terms of position and magnitude. The main parameters (magnitudes, diameters) of the four targets and their calibrators are listed in Table~\ref{tab:stars}. Because some observations were obtained during commissioning, nearby calibrators were not always available. In these cases (e.g., on the night of October 28), we carefully checked on long calibration sequences that the instrumental transfer function does not depend on the target position, which was always the case.

\begin{table}[t]
\caption{Main parameters of our four targets and their calibrators.}
\centering
\begin{tabular}{c c c c c}
\hline\hline
Star & Type & $H$ & $\theta_{\rm LD}$ & Ref. \\
\hline
Fomalhaut & A3V & $1.03$ & $2.223\pm0.022$ & 1 \\
HD\,209688 & K3III & $1.58$ & $2.71\pm0.030$ & 2 \\
HD\,219784 & G8III & $2.01$ & $2.13\pm0.025$ & 2 \\
\hline
tau Cet & G8V & $1.72$ & $2.015\pm0.011$ & 3 \\
HD\,8705 & K2III & $2.17$ & $2.06\pm0.040$ & 2 \\
HD\,14376 & K5III & $2.92$ & $1.457\pm0.019$ & 4 \\
HD\,15694 & K3III & $2.61$ & $1.766\pm0.023$ & 4 \\
HD\,16526 & K4III & $3.12$ & $1.389\pm0.017$ & 4 \\
\hline
del Aqr & A3V & $3.14$ & $0.825\pm0.012$ & 5\tablefootmark{a} \\
HD\,221745 & K4III & $3.06$ & $1.426\pm0.019$ & 4 \\
gam~Tuc & F4V & $3.00$ & $0.998\pm0.070$ & 6 \\
\hline
Regulus & B7V & $1.57$ & $1.47\pm0.20$ & 7\tablefootmark{b} \\
HD\,45018 & K5III & $2.00$ & $2.50\pm0.041$ & 2 \\
HD\,83425 & K3III & $1.68$ & $2.64\pm0.028$ & 2 \\
HD\,82381 & K2III & $2.07$ & $2.16\pm0.026$ & 2 \\
\hline
\end{tabular}
\tablefoot{References: (1) \citet{Absil09}, (2) \citet{Borde02}, (3) \citet{DiFolco07}, (4) \citet{Merand05}, (5) \citet{Kervella04}, (6) \citet{Lafrasse10}, (7) \citet{McAlister05}. \\ \tablefoottext{a}{Diameter based on surface-brightness relationships, using magnitudes $V=3.269\pm0.01$ and $K=3.078\pm0.02$ \citep{Bouchet91}.} \\ \tablefoottext{b}{The uncertainty takes into account the stellar oblateness.}}
\label{tab:stars}
\end{table}

Various baselines were used during the observations, with ground separations ranging from 16\,m (D0-E0, E0-G0) to 128\,m (A0-K0). The mean ground baseline lengths of the four configurations used during the reported observations are of 44\,m (E0-G0-H0-I1), 53\,m (D0-E0-H0-I1), 63\,m (D0-G1-H0-I1), and 86\,m (A0-G1-K0-I1), respectively. The observations are divided into observing blocks (OBs), corresponding to the lines of Table~\ref{tab:obs}. Each OB is composed of several files, which consist in 100 successive fringe measurements and typically last 1\,min. The total amount of time spent on each of the four targets, including calibration observations, amounts to about 4h (Fomalhaut), 3h (tau Cet), 30~min (del Aqr), and 1h30 (Regulus), respectively, for a total number of calibrated OBs of 7 (Fomalhaut), 6 (tau Cet), 1 (del Aqr) and 3 (Regulus). The observing conditions generally ranged from good (seeing of $0\farcs8$, $\tau_0=5$\,msec) to fair (seeing of $1\farcs2$, $\tau_0=2$\,msec), with the exception of the observations of 2010/11/28 on Regulus, where the conditions were poor (seeing of $1\farcs4$, $\tau_0=1.5$\,msec).

The interferometric concept and the data reduction strategy of PIONIER are directly inspired from the IONIC-3 experiment \citep{Berger01}. The interested reader can find a detailed description of the instrument and its performance in \citet{LeBouquin11}. In brief, PIONIER relies on a pairwise co-axial combination embedded inside an integrated optics component. Before the flux is injected into the chip, the four beams are modulated at non redundant velocities $(-3v,-v,+v,+3v)$ to generate the six fringe signals. The length of the shortest scan is generally set to $80\,\mu$m. For bright stars, $v$ is imposed by the detector frame rate and the need to correctly sample the fringe at the highest frequency $f_{max}=6v/\lambda$. In the observations presented here, the velocity was set to $v=70\,\mu$m/s, which corresponds to fringes at $80$, $160$, and $240\,$Hz. This is just fast enough to freeze the turbulent optical path delay (OPD) under median atmospheric conditions. The outputs of the chip are dispersed over seven spectral channels across the H-band, providing low ($R=35$) spectral resolution. During the observation, each scan is processed by a quick-look algorithm that also implements slow group tracking.

All the data have been reduced with the $\texttt{pndrs}$ package described in \citet{LeBouquin11}. For each file and each spectral channel, the $\texttt{pndrs}$ package provides six visibility and four closure phase measurements. The statistical uncertainties are estimated by the dispersion over the 100 scans contained in each file, and typically range from 0.25 to 3~degrees for the closure phase, depending on the target brightness and atmospheric conditions. The statistical uncertainty includes contributions from the classical detection noises (detector and photon noise), as well as from the atmospheric OPD fluctuations faster than the scan rates. The instrumental and atmospheric contributions to the visibilities and closure phases (the so-called transfer function) is monitored by interleaving the observation of science stars with calibration stars. Some nights show an extremely stable transfer function, down to $0.1\,$deg, while other nights require fitting the transfer function by a drift, indicating that we are facing some non-stationary biases. This aspect requires additional commissioning before being fully understood.


\section{Data analysis strategy} \label{sec:analysis}

In this section, we discuss how the PIONIER calibrated data set is used to search for faint companions, compute confidence levels for tentative detections, and derive upper limits to the presence of off-axis companions in case of non-detections. Our analysis is based solely on the calibrated closure phases, which have two significant advantages over (differential) visibilities and (differential) phases. On the one hand, at first order, closure phases are not affected by phase disturbances ahead of the beam combiner \citep[see e.g.,][]{Monnier03}. Therefore, closure phases are not affected by atmospheric turbulence and other instrumental effects such as mechanical vibrations, which strongly reduce the accuracy of other observable quantities. On the other hand, closure phase are especially sensitive to faint off-axis companions, as they differ from zero only for non point-symmetric targets. One of the associated benefits is that the detection of a faint companion does not depend on the knowledge of the parent star's photospheric parameters, provided that it is point-symmetric; however, poor knowledge of the primary stellar diameter may lead to larger uncertainty on the photometry of a potential companion.

Our companion search strategy is based on the computation of the $\chi^2$ distance between our data set and several models where a faint companion is added to a user-specified photospheric model for the central star. The $\chi^2$ goodness-of-fit is computed for a series of angular separations along east and north between the primary star and its companion. The first step in this analysis is therefore to define the region around the star in which companions will be searched for.

	\subsection{Determining the search region}

There are three limitations to the maximum angular separation at which a companion can be searched for in the PIONIER data. The first one is related to the use of single-mode fibres in the instrument. The coupling efficiency of point-like sources into a single-mode fibre depends on the angular separation between the source and the optical axis. The transmission profile, which results from the overlap integral between the turbulent image of a point-like source and the fundamental mode of the fibre can be approximated by a Gaussian profile, whose full width at half maximum (FWHM) depends on the fibre parameters, the telescope diameter, the observing wavelength, and the strength of the atmospheric turbulence. In the case of PIONIER, working with the 1.8\,m ATs under typical atmospheric conditions (seeing of $0\farcs8$ in the visible), the FWHM of the Gaussian profile is about 400\,mas. The transmitted flux of off-axis companions located at angular separations larger than 200\,mas from the central will therefore be attenuated by more than 50\%.

A second limitation to the angular distance of detectable companions comes from the separation of the fringe packets associated to the two stars in terms of optical path delay (OPD). To properly detect a potential companion, we ask the fringe packet separation to be less than half the size of the smallest scan, which amounts to $40\,\mu$m for the baselines with the slowest scanning speed \citep{LeBouquin11}. The OPD separation of the fringe packets is directly related to the angular separation $\Delta\theta$ of the two objects in the sky: $\Delta{\rm OPD} = B \, \Delta\theta \, \cos\theta$, with $B \, \cos \theta$ the projected baseline length. Using a mean projected baseline of about 40\,m for the compact configuration (E0-G0-H0-I1, used for most targets) and taking $\Delta{\rm OPD}_{\rm max}=20\,\mu$m, we end up with $\Delta\theta_{\rm max} \simeq 100$\,mas. A fringe packet separation smaller than $20\,\mu$m also ensures that the two fringe packets are (at least partially) superposed in all cases, as the size of the fringe packet is $\lambda^2/\Delta\lambda \simeq 63\,\mu$m when the H-band signal is dispersed onto seven spectral channels.

The last limitation to the size of the search region comes from the spectral sampling of the closure phase signal. Because the angular resolution of the array is proportional to the observing wavelength, the closure phase signature of an off-axis companion oscillates around zero as a function of wavelength. These oscillations must be properly sampled by the spectral resolution of the instrument in order to produce a unique (non-aliased) solution in the fitting process. As discussed by \citet{Absil10a}, the period in the closure phase signal, which is roughly given by $P_{\lambda} = \lambda^2 / (B\,\Delta\theta-\lambda)$, must be larger than four times the spectral channel size. In the present case, considering a mean sky-projected baseline $B=40$\,m for the compact configuration (the most frequently used in our observations) and taking the channel width of $0.045\,\mu$m in the PIONIER dispersed data into account, this constraint translates into $\Delta\theta_{\max} \simeq 72$\,mas. This does not prevent detecting companions located further away, but these companions can possibly produce aliases within the inner 72\,mas region and therefore lead to an ambiguous position.

Taking these three limitations into account, we restrict our search region to a radius of about 100\,mas around the target stars. Detections within this region should be mostly robust and unambiguous. It must, however, be noted that companions located further away (within about 500\,mas from the central star) could lead to ``false'' detections within our 100\,mas search region. The presence of such companions would, however, generally be known from AO-assisted (coronagraphic) observations on 10-m class telescopes (see Sect.~\ref{sec:intro}).

		\subsection{Searching for companions} \label{sub:search}

\begin{figure*}[!t]
\centering
\resizebox{\hsize}{!}{\includegraphics{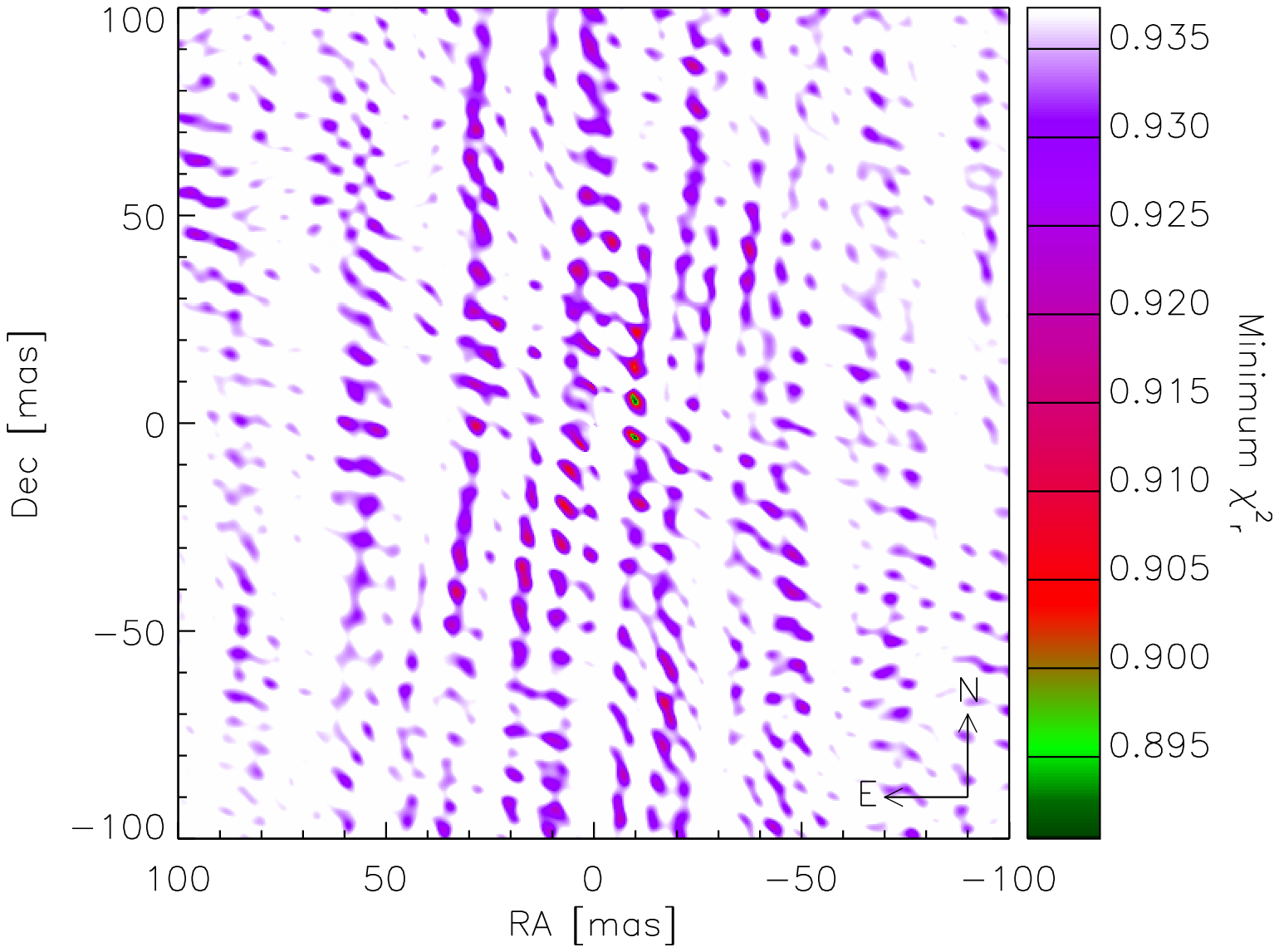} \includegraphics{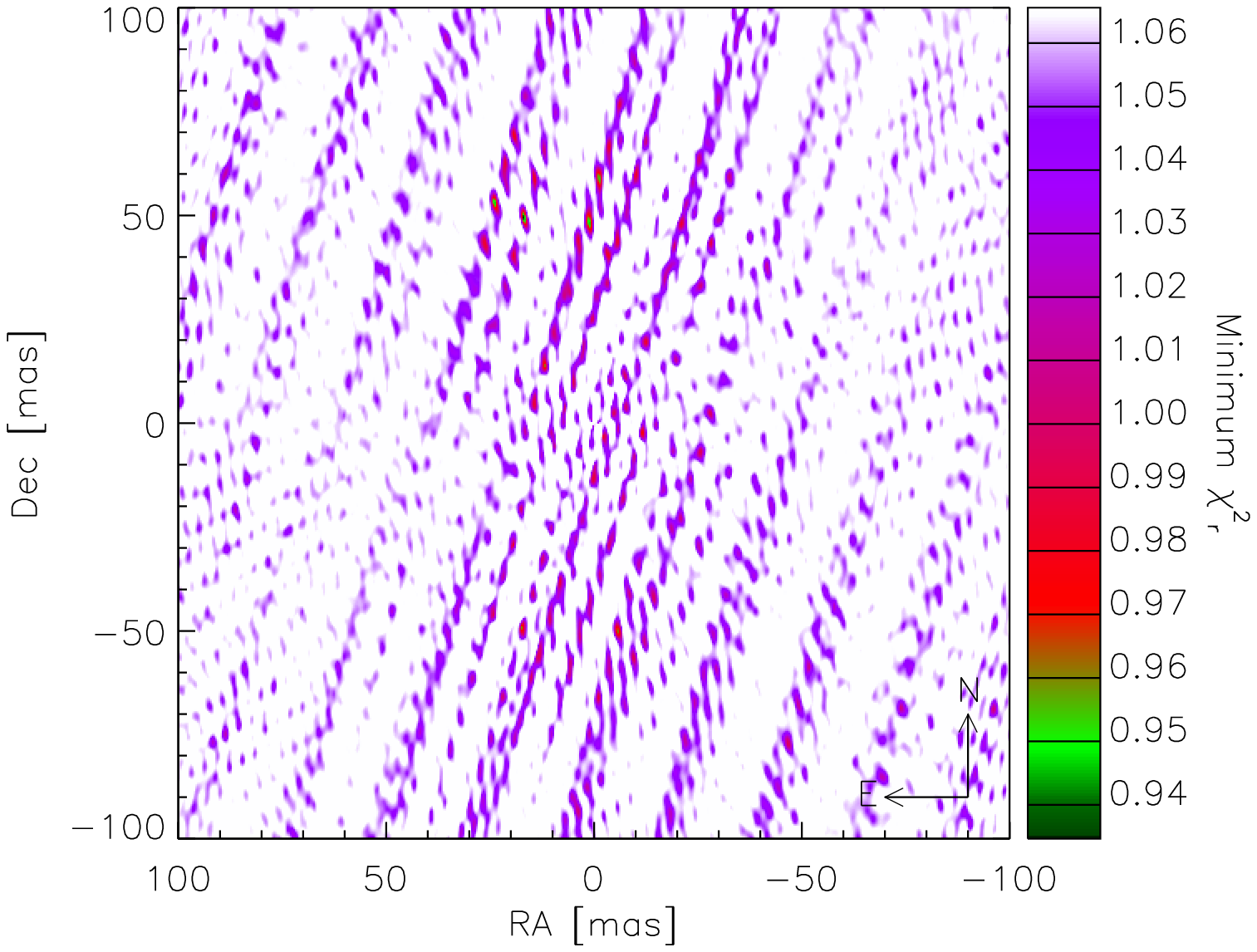}}
\resizebox{\hsize}{!}{\includegraphics{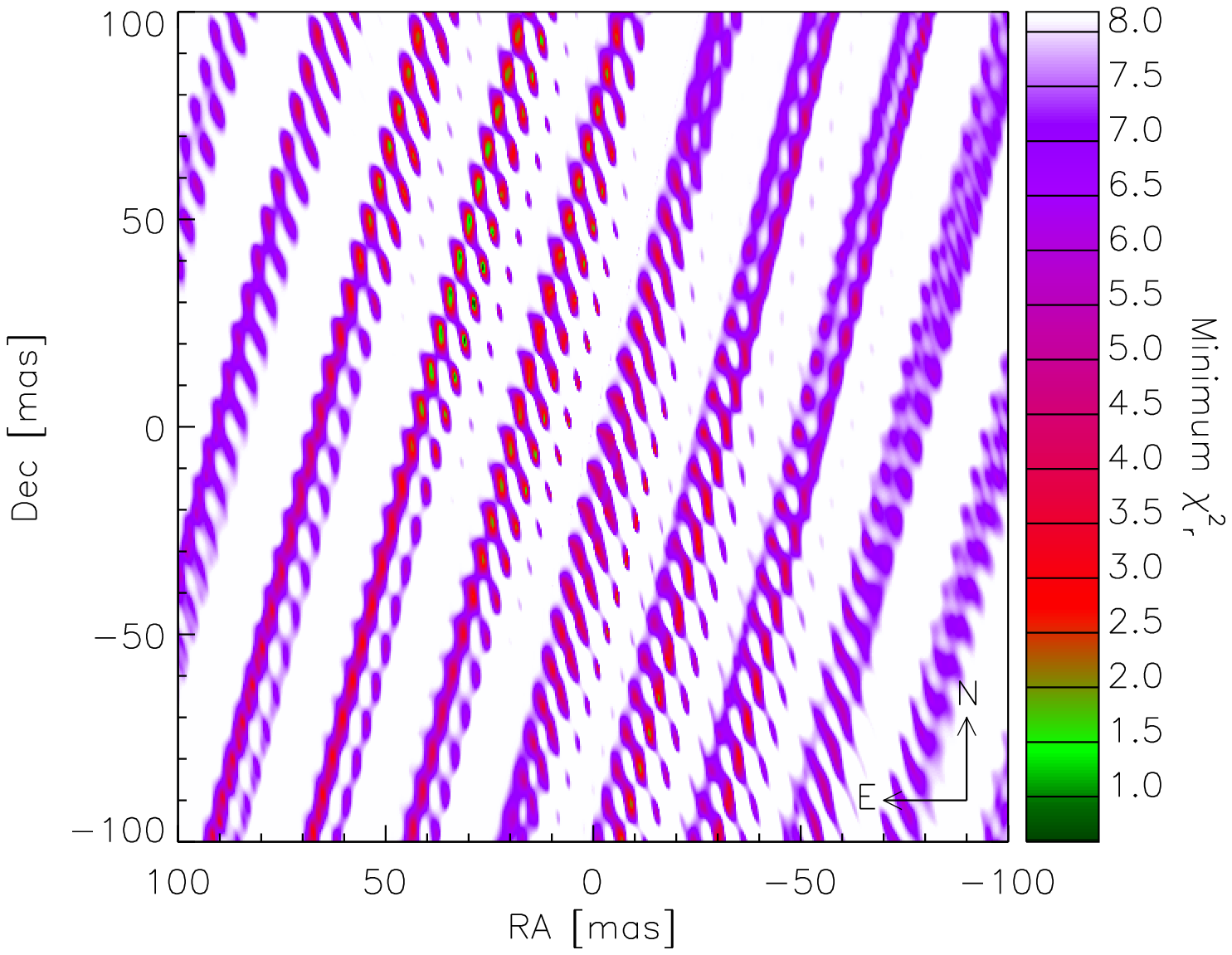} \includegraphics{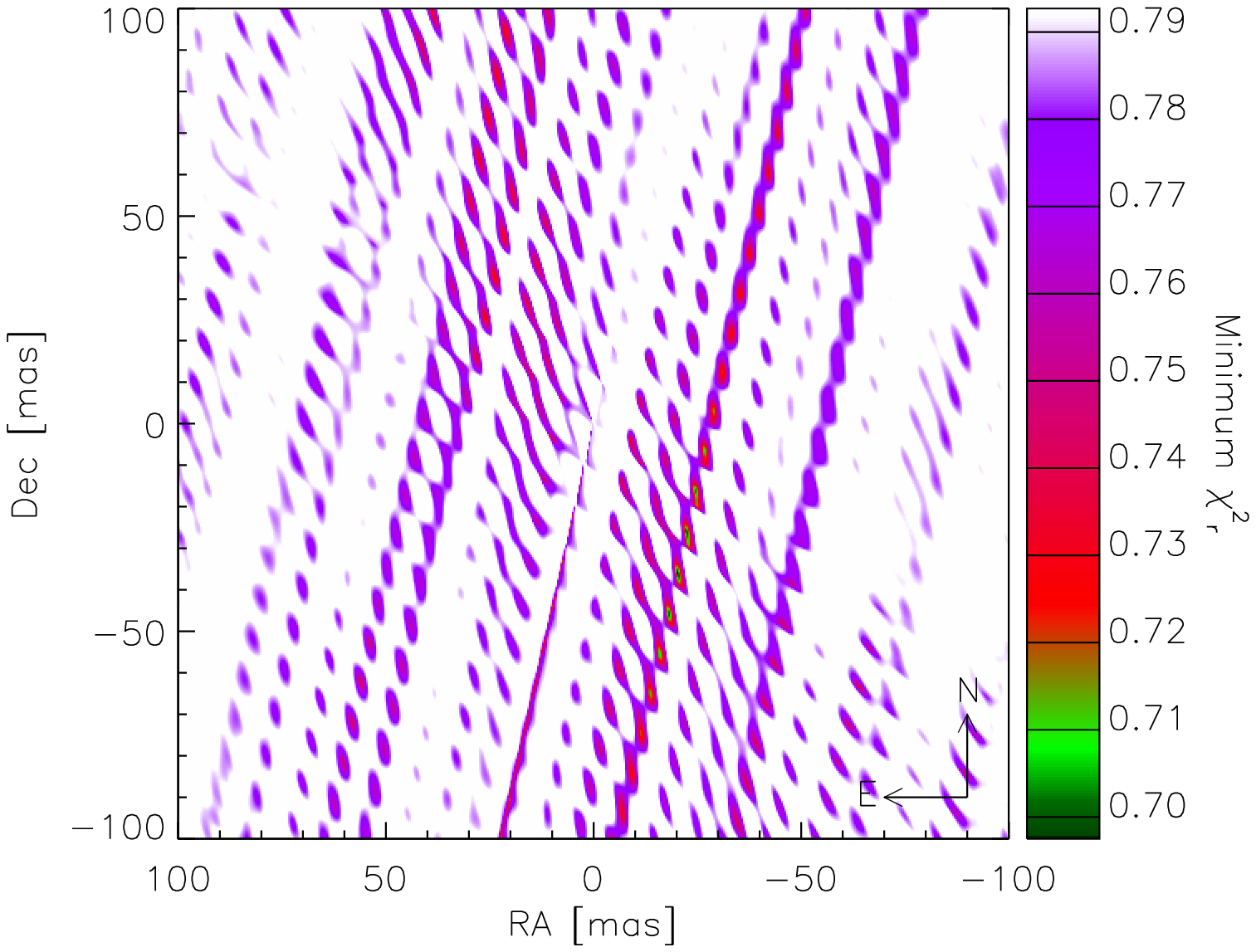}}
\caption{Two-dimensional representation of the $\chi_r^2$ cubes, showing the minimum $\chi^2_r$ obtained at each point of the search region when exploring the flux ratio of our binary star models in an attempt to reproduce the Fomalhaut (\textit{top left}), tau Cet (\textit{top right}), del Aqr (\textit{bottom left}), and Regulus (\textit{bottom right}) data sets. These maps are computed before any renormalisation of the $\chi^2_r$ cube.}
\label{fig:chi2maps}
\end{figure*}

Our search strategy consists in building a series of binary models with off-axis companions at all possible locations $(x,y)$ within the search region and with various flux ratios $r$, starting at $r=0$ to include the single-star model. Our binary models feature a realistic description of the primary stellar photosphere, based on interferometric measurements where available or on surface-brightness relationships \citep{Kervella04}. They include the effects of limb-darkening and of a possible elongation of the photosphere due to rapid stellar rotation, as described in \citet{Absil08}. The off-axis companions are supposed to be fully unresolved by the interferometer (i.e., point-like) and are thus only parameterised by their flux ratio $r$ with respect to the primary star. For each individual model $(x,y,r)$, we compute the associated closure phases at the relevant observing dates and wavelengths, and derive the $\chi^2$ goodness-of-fit to the whole data set. This quantity is divided by the number of degrees of freedom $\nu$ of our $\chi^2$ variable to obtain a cube of reduced $\chi^2$, noted $\chi^2_r(x,y,r)$. The search strategy is then based fully on this cube and on the assumption that our data set follows purely Gaussian statistics. A 2D illustration of the $\chi^2_r$ cube for our four targets is given in Fig.~\ref{fig:chi2maps}, where the minimum $\chi^2_r$ as a function of flux ratio has been selected at each position. In these maps, the maximum value in the scale always corresponds to the single-star model.

The following step is to search for the global minimum of the $\chi^2_r$ cube and to renormalise the cube so that $\chi^2_r=1$ for the global best-fit binary model. In doing so, we implicitly assume that this best-fit model actually corresponds to the true nature of the observed source. Based on the renormalised $\chi^2_r$, we compute the probability $P_0$ for a random variable in the $\chi^2_r$ distribution to be equal to or larger than the $\chi^2_r$ associated to the single-star model ($r=0$):
\begin{equation}
P_0 = 1 - {\rm CDF}_{\nu}\left(\frac{\nu\chi^2_r(r=0)}{\min_{x,y,r}(\chi^2_r(x,y,r))}\right) \, ,
\label{eq:probazero}
\end{equation}
where ${\rm CDF}_{\nu}$ is the $\chi^2$ cumulative probability distribution function with $\nu$ degrees of freedom. If $P_0$ is below a pre-defined threshold, we can reject the single-star model and the best-fit binary solution is considered as significant. The threshold is generally fixed at a $3\sigma$ level, i.e., at a probability of 0.27\%. The error bars on the best-fit flux ratio and position are then determined by inspecting the dependence of the $\chi^2_r$ around the global minimum with respect to each of the three parameters. In practice, we fix one parameter, search the minimum $\chi^2_r$ with respect to the two other parameters, and repeat the same procedure for several values of the first parameter around the minimum. The resulting $\chi^2_r$ is then translated into a probability, and the error bars are defined as the 68\% confidence interval around the minimum for each individual parameter.

Finally, we convert the 2D map of the minimum renormalised $\chi^2_r$ into a 2D probability map, following the same procedure as in Eq.~\ref{eq:probazero}: 
\begin{equation}
P_{\rm map}(x,y) = 1 - {\rm CDF}_{\nu}\left(\frac{\nu\min_r(\chi^2_r(x,y,r)}{\min_{\alpha,\beta,\rho}(\chi^2_r(\alpha,\beta,\rho))}\right) \, .
\label{eq:probamap}
\end{equation}
A visual inspection of this map is performed to check for multiple solutions. We also inspect the closure phase data and the best-fit closure phases to look for possible outliers or systematics effects on our data set, which could produce artefacts in our search results.

		\subsection{Deriving sensitivity limits} \label{sub:sens}

In case no companion is detected, we derive sensitivity limits as follows. The $\chi^2_r$ cube is first renormalised so that $\chi^2_r=1$ for the single-star model. This corresponds to assuming that there is actually no companion around the observed star. The renormalised $\chi^2_r$ cube is then converted into a probability cube:
\begin{equation}
P_{\rm cube}(x,y,r) = 1 - {\rm CDF}_{\nu}\left(\frac{\nu\chi^2_r(x,y,r)}{\chi^2_r(r=0))}\right) \, .
\label{eq:probacube}
\end{equation}
For each position $(x,y)$, we start at $r=0$ and increase the flux ratio of the off-axis companion until the associated probability becomes lower than a pre-defined threshold, usually 0.27\% for a $3\sigma$ significance level. If such a companion had been present in the data, the single-star model would have been inconsistent with the data at $3\sigma$ or more, and a detection would have been reported. This flux ratio is the lowest that could have been detected at that particular position. By repeating this procedure at all positions, we create a sensitivity map giving the $3\sigma$ upper limit on the flux ratio of potential companions:
\begin{equation}
S_{\rm map}(x,y) = r \, , \: {\rm so \: that} \: P_{\rm cube}(x,y,r)=0.27\% \, .
\label{eq:sensmap}
\end{equation}
This map can then be used to produce (cumulated) histograms of the sensitivity on the search region, and define the sensitivity limit for various completeness levels.


\section{Results}

    \subsection{Long integrations on Fomalhaut and tau Cet} \label{sub:longint}

\begin{figure*}[t]
\centering
\resizebox{\hsize}{!}{\includegraphics{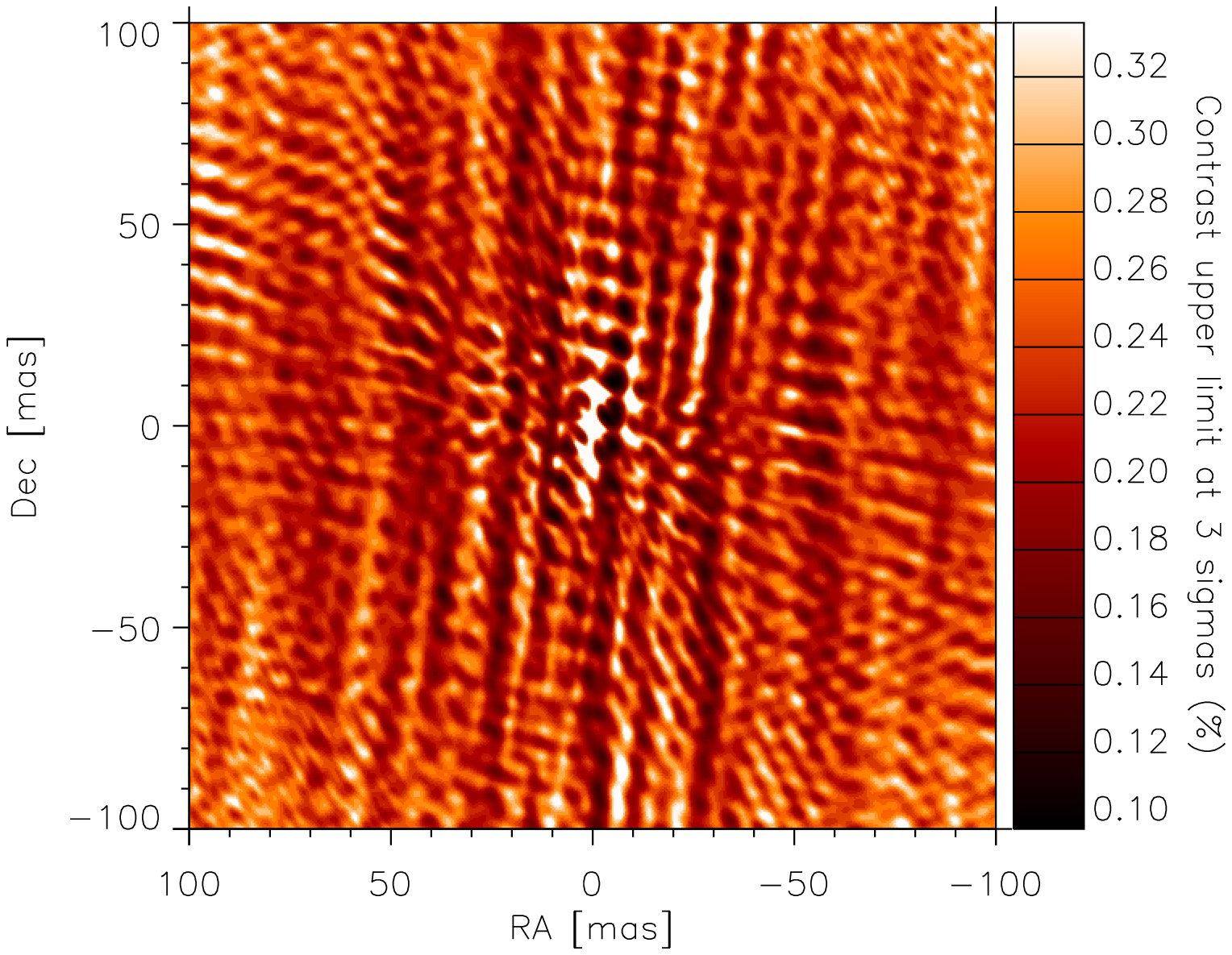} \includegraphics{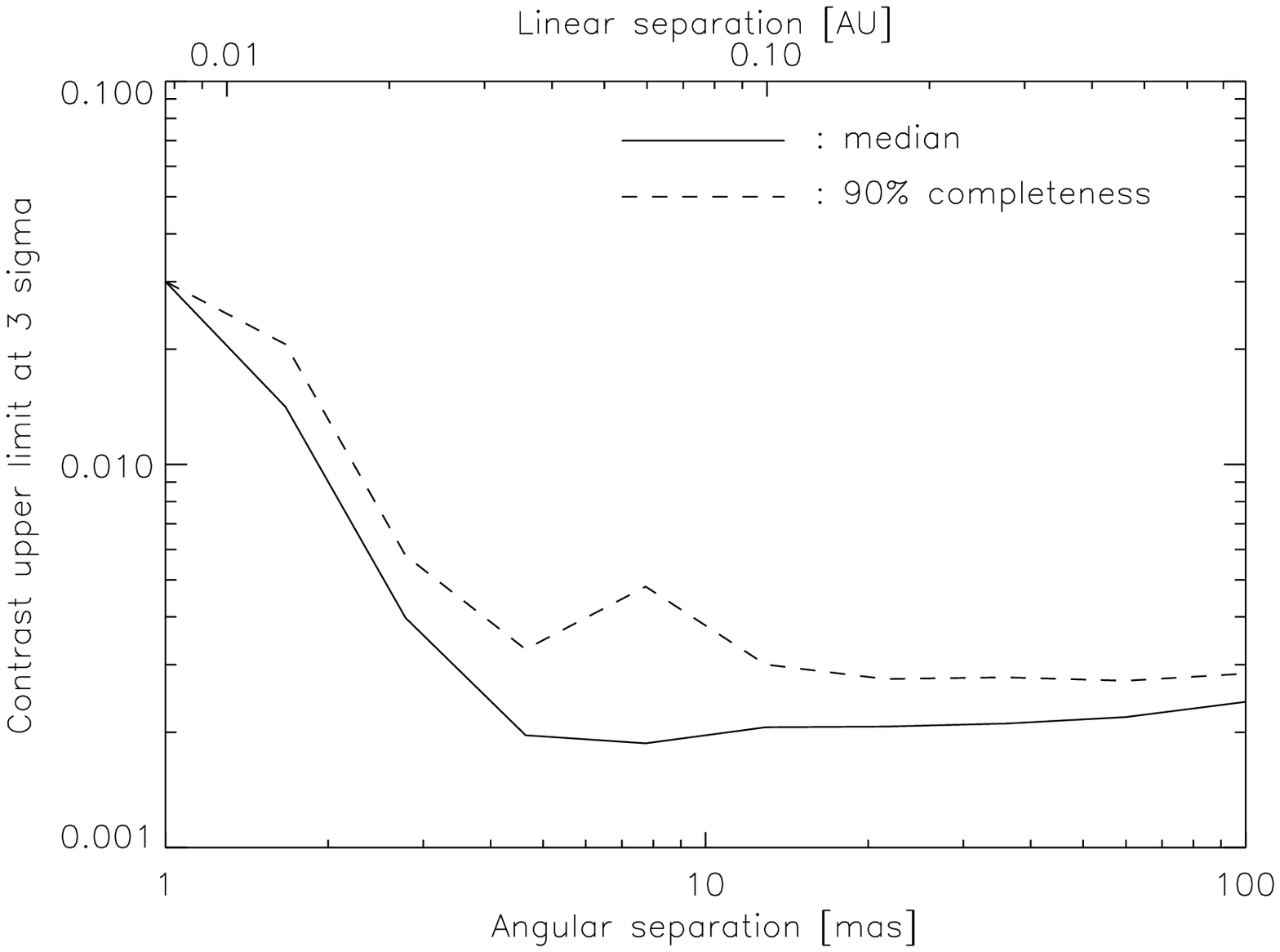}}
\resizebox{\hsize}{!}{\includegraphics{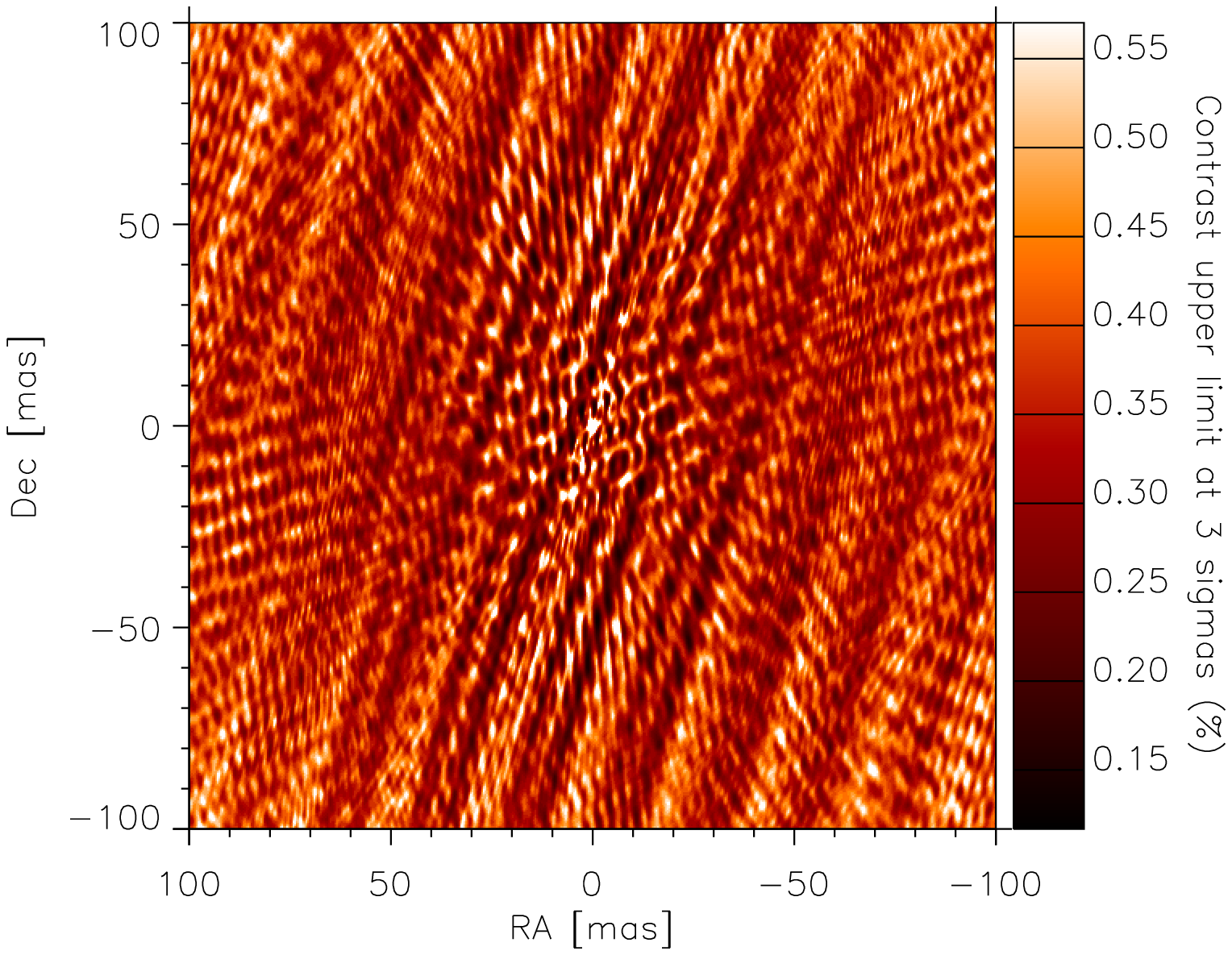} \includegraphics{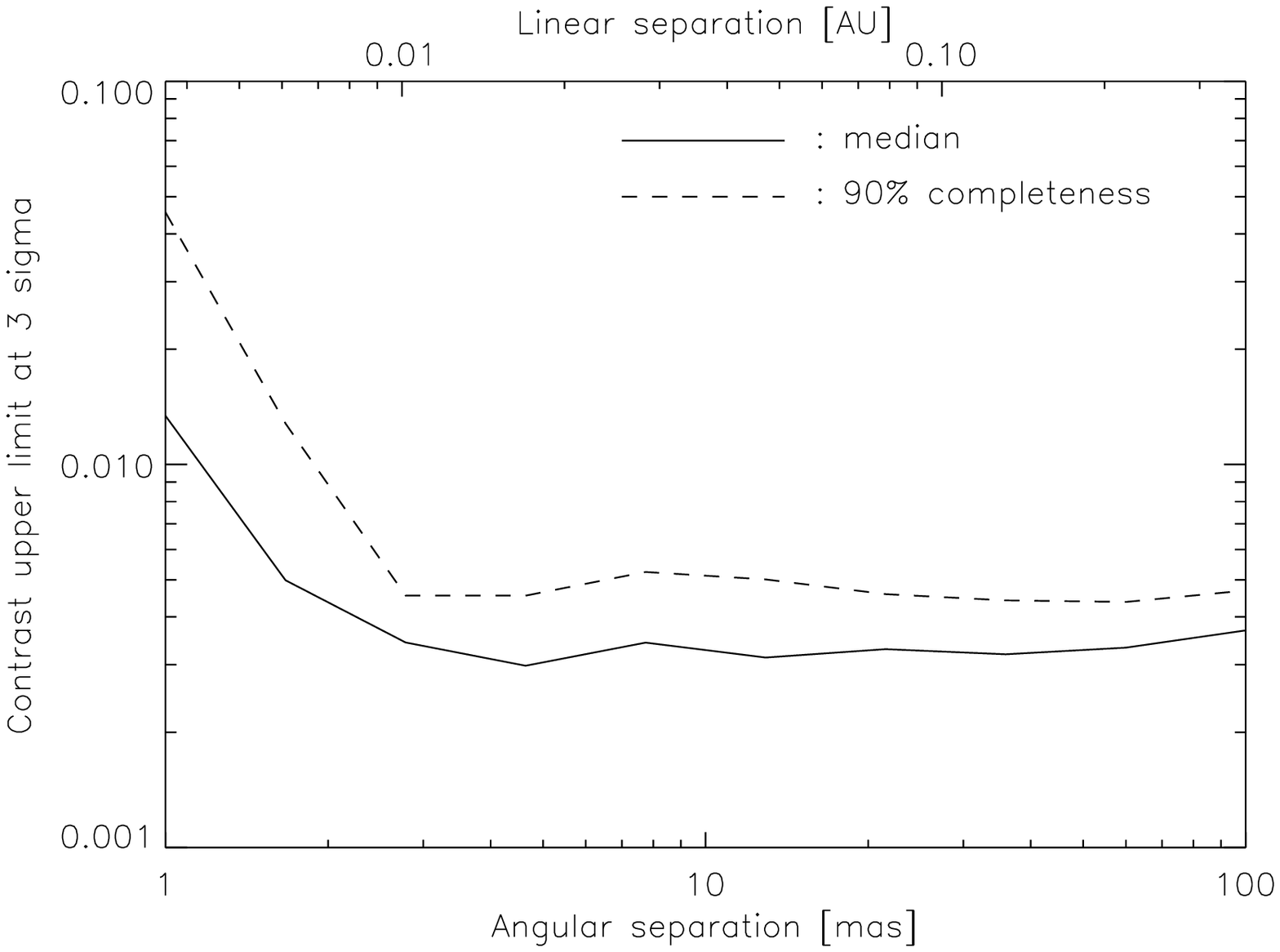}}
\resizebox{\hsize}{!}{\includegraphics{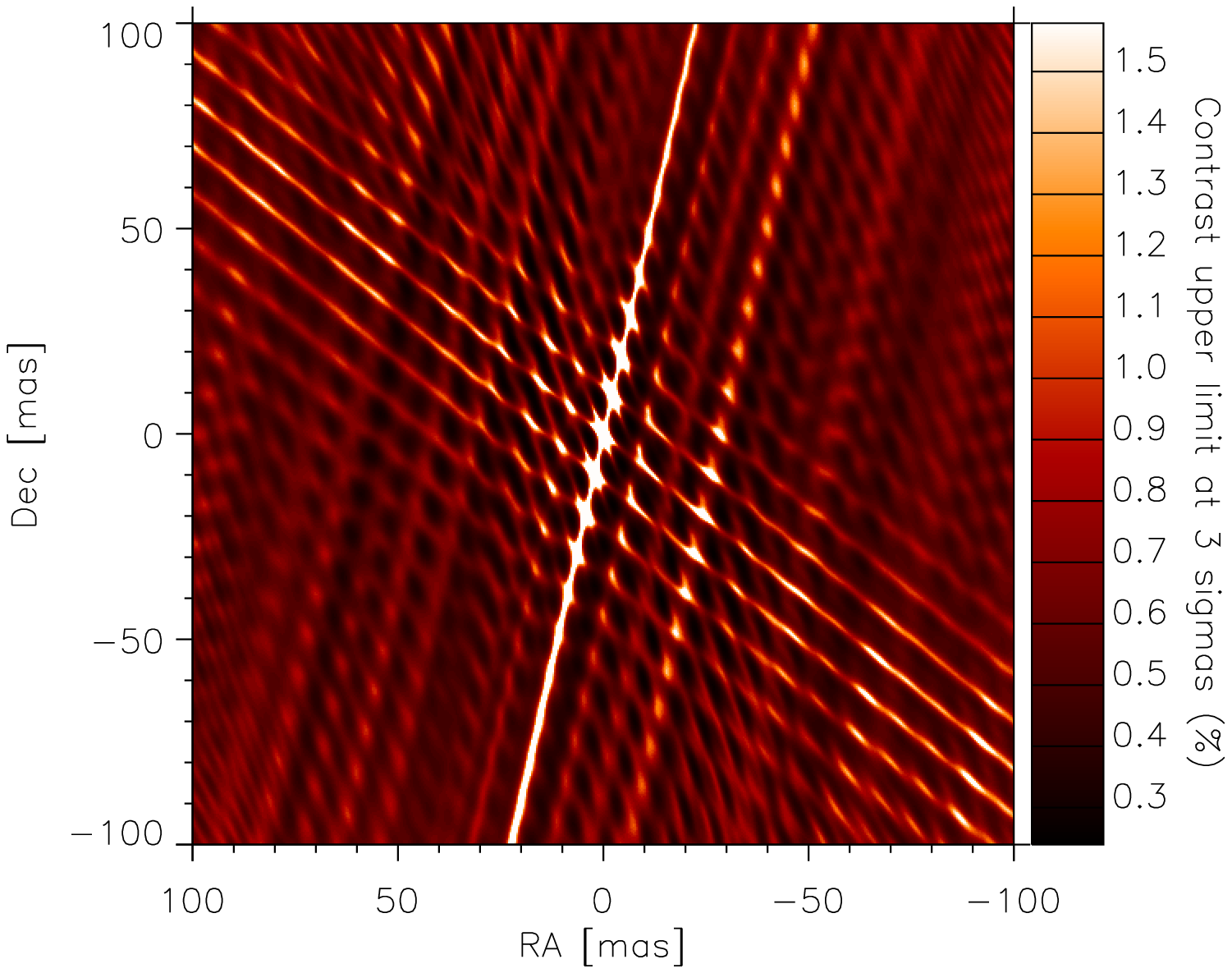} \includegraphics{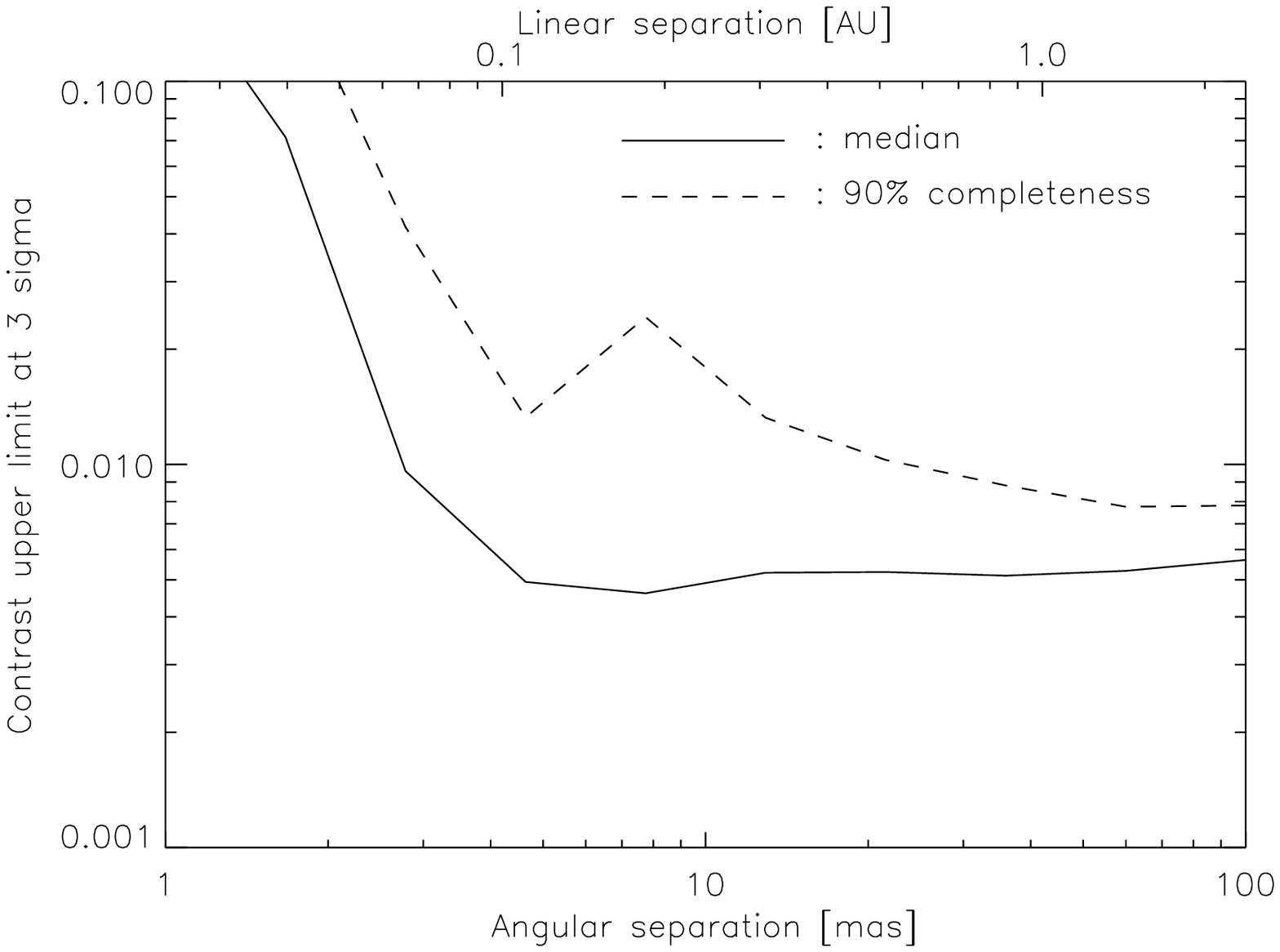}}
\caption{\textit{Left.} Map of the $3\sigma$ upper limit on the flux ratio of companions around Fomalhaut (top), tau Cet (middle) and Regulus (bottom). The uppermost 1\% values have been clipped to reduce the colour scale range. \textit{Right.} Associated sensitivity as a function of angular distance, for two completeness levels (50\% or 90\%). The 50\% completeness level corresponds to the median $3\sigma$ sensitivity.}
\label{fig:sensitivity}
\end{figure*}

Deep integrations, of about 3h and 4h respectively (including calibration time), have been obtained on Fomalhaut and tau Cet. We used the whole data set for each star to compute the minimum $\chi^2_r$ maps, which are displayed in Fig.~\ref{fig:chi2maps}, before any renormalisation. The number of degrees of freedom for the $\chi^2$ variable is respectively 1154 and 629. With such large data sets, it is expected that any potential systematic error(s) are mostly averaged out, leaving a clean Gaussian statistics. Indeed, it must be noted that the $\chi^2_r$ values are close to unity in both cases, which confirms that error bars have been properly evaluated by the data reduction software and which suggests that systematic errors do not dominate the overall error budget. A quick inspection of the best $\chi^2_r$ maps shows a relatively high uniformity of the best $\chi^2_r$ (narrow range of values), and the absence of particularly outstanding minima.

To properly assess the presence of significant minima in the maps, we renormalise the $\chi^2_r$ cube so that $\chi^2_r=1$ for the best-fit model, as described in Section~\ref{sub:search}. The single-star model is then associated with a $\chi^2_r$ of 1.05 for Fomalhaut and 1.14 for tau Cet. The corresponding probability for the single-star model to reproduce the data is 10\% (or $1.6\sigma$) and 0.7\% (or $2.7\sigma$), respectively. The minima of the $\chi^2_r$ maps can therefore not be considered as significant, and we conclude that no firm detection is reported around these two stars.

Following Eq.~\ref{eq:sensmap}, we derive the $3\sigma$ upper limits on the flux of off-axis companions within the search region around both stars. The resulting sensitivity maps are illustrated in Fig.~\ref{fig:sensitivity} (top and middle left). They show sensitivity limits in the range 0.10\%--0.34\% for Fomalhaut and 0.11\%--0.57\% in the case of tau Cet, suggesting that companions as faint as 1:1000 of the stellar flux would be detectable at some particular positions within the search region. The cumulated histogram of the sensitivity level across the whole search region is illustrated by the solid curves in the left-hand side plots of Fig.~\ref{fig:blind_test}. The median and 90\% percentile sensitivity limits are $2.3\times10^{-3}$ and $2.8\times10^{-3}$ (Fomalhaut) and $3.5\times10^{-3}$ and $4.5\times10^{-3}$ (tau Cet), respectively. Another, more conventional way to represent the sensitivity of our observations is to consider concentric annuli, in which cumulated sensitivity histograms can be inspected to search for, e.g., median and 90\% completeness levels. This is illustrated on the right-hand side of Fig.~\ref{fig:sensitivity} (top and middle). The full dynamic range of PIONIER is achieved as close as 5\,mas for Fomalhaut and 3\,mas for tau Cet, and is maintained all the way to the edge of the search zone (100\,mas), where the effect of the Gaussian profile of the fibre transmission starts to be noticeable.

    \subsection{Short integrations on del Aqr and Regulus} \label{sub:delaqrregulus}

\begin{figure}[!t]
\centering
\resizebox{\hsize}{!}{\includegraphics{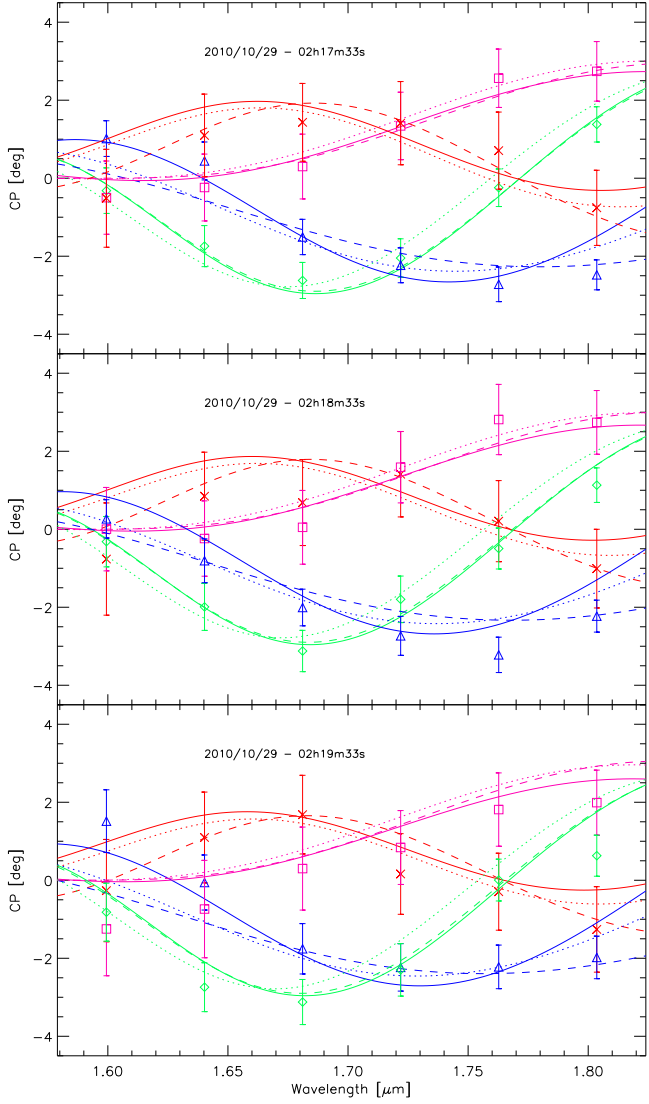}}
\caption{Calibrated closure phases for the four triangles and the three MJDs of the del Aqr data set. The closure phases of the three best-fit binary models are represented by solid, dotted, and dashed lines.}
\label{fig:fitdelaqr}
\end{figure}

\begin{figure}[!t]
\centering
\resizebox{\hsize}{!}{\includegraphics{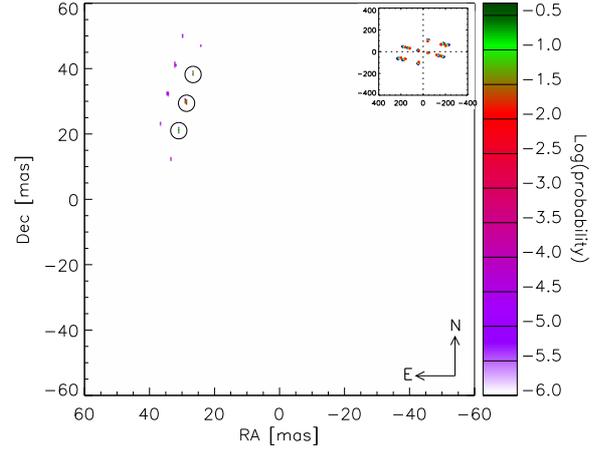}}
\caption{Probability (Eq.~\ref{eq:probamap}) for the best-fit binary models to reproduce the data set on a $60\times60$\,mas search region around del Aqr. The highest three maxima are identified by circles. The $u,v$ coverage associated to this data set is represented in the upper right inset, including wavelength dependence (using colours from blue to red), with the axes graduated in cycles per arcsec.}
\label{fig:probadelaqr}
\end{figure}

To illustrate the capabilities of short integrations (``snapshots'') in terms of companion detection limits, we have selected two representative targets observed during the first scientific runs of PIONIER: del Aqr and Regulus. The former was observed on only one occasion (see Table~\ref{tab:obs}). The data set shows very obvious non-zero closure phases (see Fig.~\ref{fig:fitdelaqr}). The associated minimum $\chi^2_r$ map (Fig.~\ref{fig:chi2maps}, bottom left) shows that the single-star model is inconsistent with the data set, giving $\chi^2_r=8.2$ before any renormalisation of the error bars. This $\chi^2_r$ increases to 14 once the error bars have been renormalised to obtain $\chi^2_r=1$ for the best-fit binary model. With 53 degrees of freedom in the $\chi^2$ distribution, the ``null hypothesis'' (absence of companion) can then be rejected at about $40\sigma$. In Fig.~\ref{fig:probadelaqr}, we show the probability map defined in Eq.~\ref{eq:probamap}. Owing to the very incomplete $u,v$ plane coverage provided by a single OB (see inset in Fig.~\ref{fig:probadelaqr}), the position of the detected off-axis object cannot be unambiguously determined. Three positions are almost as likely (minimum $\chi^2_r$ ranging from 1.00 to 1.16, i.e., at less than $1\sigma$ from each other), and are respectively located at (E: $31.0\pm0.10$\,mas, N: $21.0\pm0.26$\,mas), (E: $28.6\pm0.18$\,mas, N: $29.4\pm0.34$\,mas), and (E: $26.6\pm0.12$\,mas, N: $38.2\pm0.27$\,mas). All other positions have probabilities below $10^{-3}$. The best-fit flux ratio, which amounts to $2.05\times10^{-2} \pm 0.16\times10^{-2}$, is compatible within $1\sigma$ for the three possible positions. The closure phases associated to the three best-fit solutions are displayed in Fig.~\ref{fig:fitdelaqr}.

In the case of Regulus, three successive OBs have been obtained (see Table~\ref{tab:obs}). This is representative of the quantity of data one would get in a large survey for companions around main sequence stars, although the number of files per OB (3 to 5) was rather low in the present case. The analysis of this data set with the $\chi^2$ method results in the absence of significant detection: once the $\chi^2_r$ cube has been renormalised as discussed in Section~\ref{sub:search}, the single-star model has $\chi^2_r=1.14$, which corresponds to a probability of $7.6\%$ (equivalent to $1.8\sigma$) to reproduce the data set, taking the 230 degrees of freedom into account. The single-star model can therefore not be rejected. We compute sensitivity limits following Section~\ref{sub:sens}, leading to the sensitivity map displayed in Fig.~\ref{fig:sensitivity} (bottom left). The bright lines within the map are due to the limited $u,v$ coverage, which leads to low sensitivities at some particular positions within the search region. The median and 90\% percentile sensitivity limits computed across the whole search region are respectively of $5.4 \times 10^{-3}$ and $7.9 \times 10^{-3}$. The corresponding sensitivity curves with respect to angular separation are shown on the right-hand side of Fig.~\ref{fig:sensitivity} (bottom), for 50\% (median) and 90\% completeness levels.


\section{Discussion}

    \subsection{Double-blind search for fake companions} \label{sub:blind}

To validate the sensitivity limits computed in Section~\ref{sub:longint} for the deep integrations, we performed a double-blind test to search for fake companions in the data sets where no companion was initially detected. One of the co-authors introduced into the calibrated data set a series of fake companions of various flux ratios, located at various positions within the search region. The range of angular separations was limited to a minimum of 5\,mas to ensure an optimum sensitivity, and to a maximum of 50\,mas to reduce computation time (extending it to 100\,mas would not have significantly changed the results). Then, the first author, unaware of the flux ratios and positions of the fake companions, performed a search for companions on the new data sets using the procedure described above. In all cases, the overall best-fit binary model was identified and the $\chi^2_r$ cube renormalised so that this best-fit model corresponds to $\chi^2_r=1$. The probability that the single-star model would reproduce the data set was then computed and converted into a number of sigmas, as in Section~\ref{sub:search}. The flux ratio and position of all best-fit models were finally compared with the input fake companion parameters.

			\subsubsection{Long integrations on Fomalhaut and tau Cet}

\begin{figure*}[t]
\centering
\resizebox{\hsize}{!}{\includegraphics{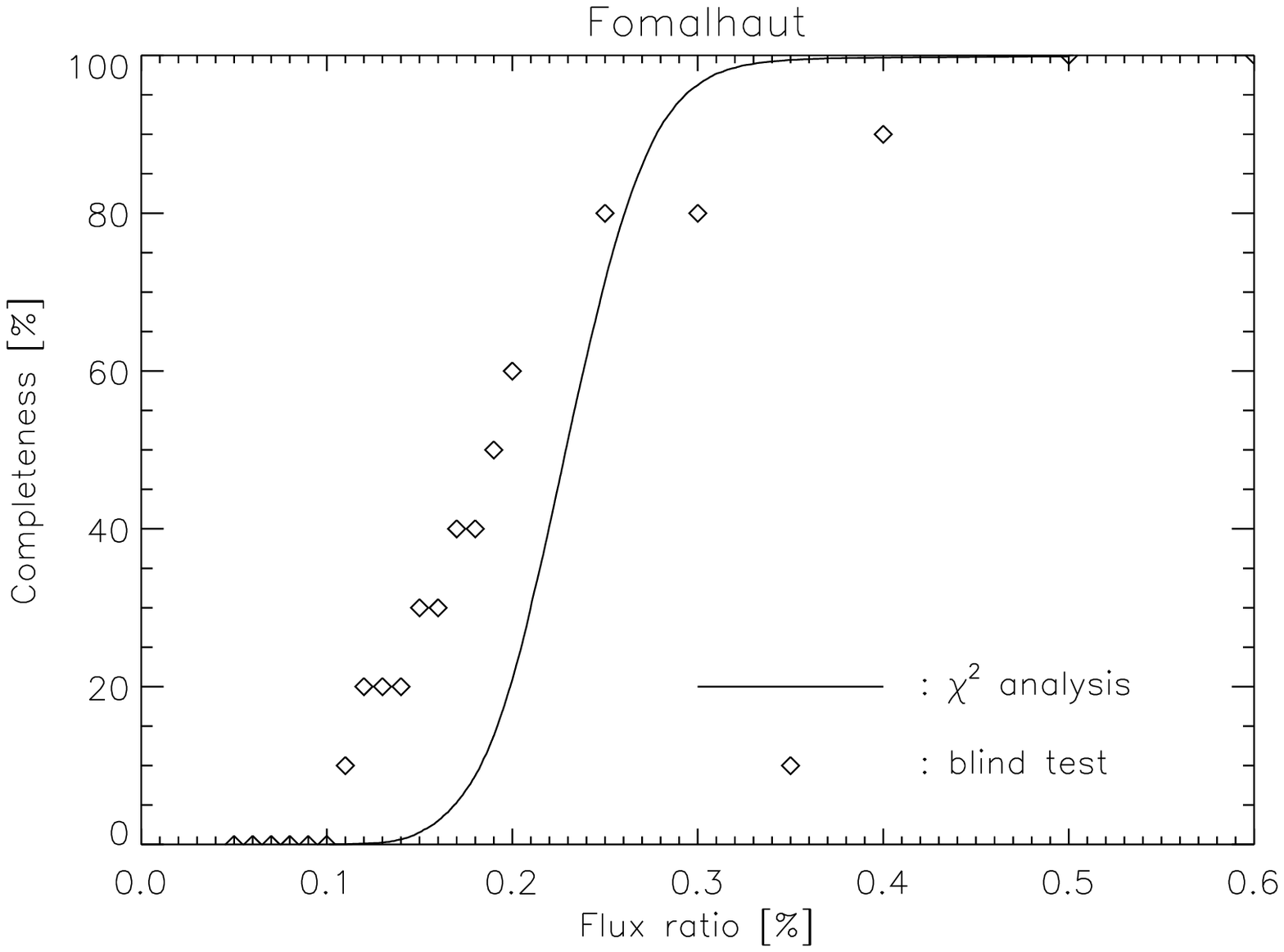} \includegraphics{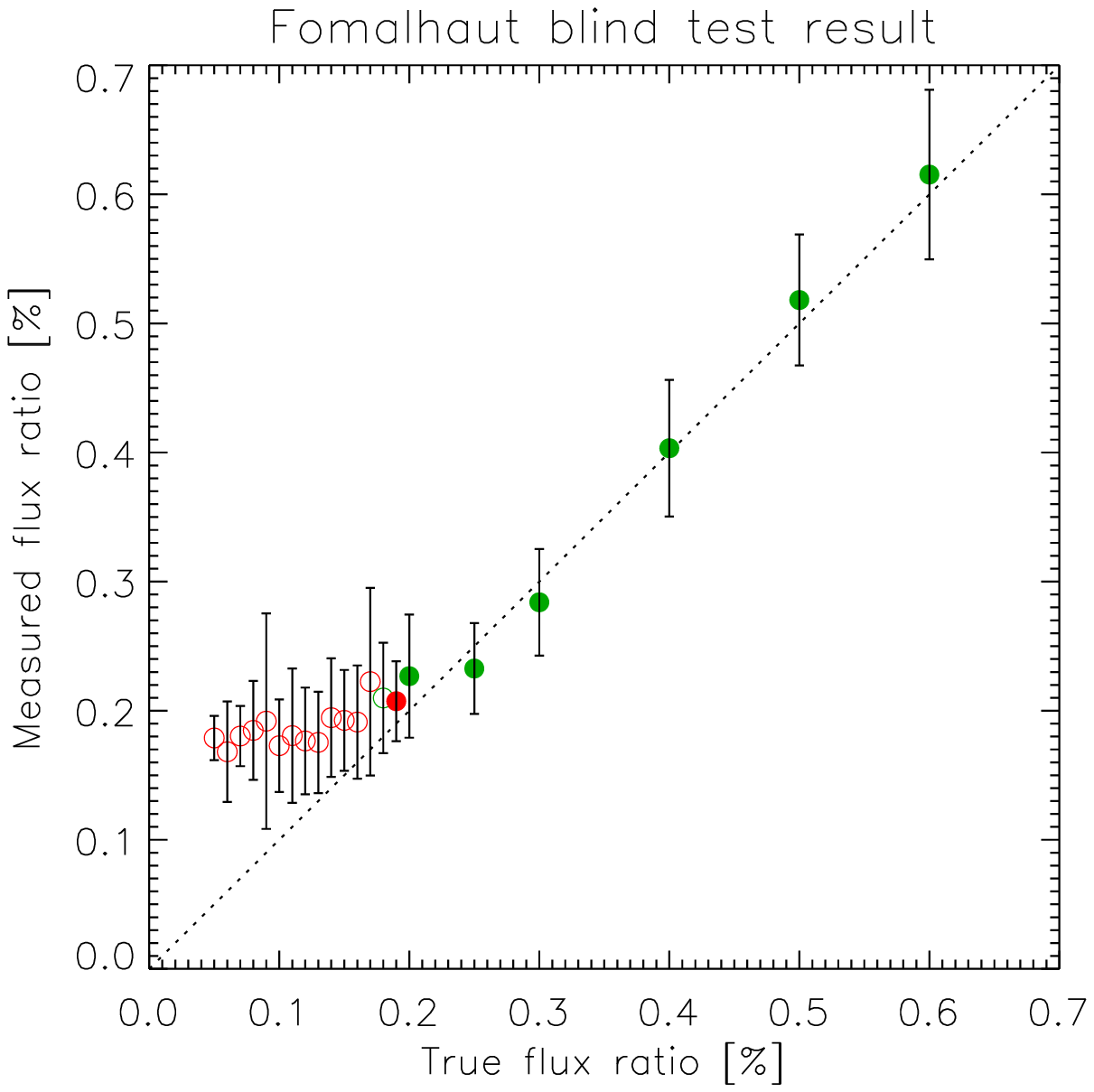}}
\resizebox{\hsize}{!}{\includegraphics{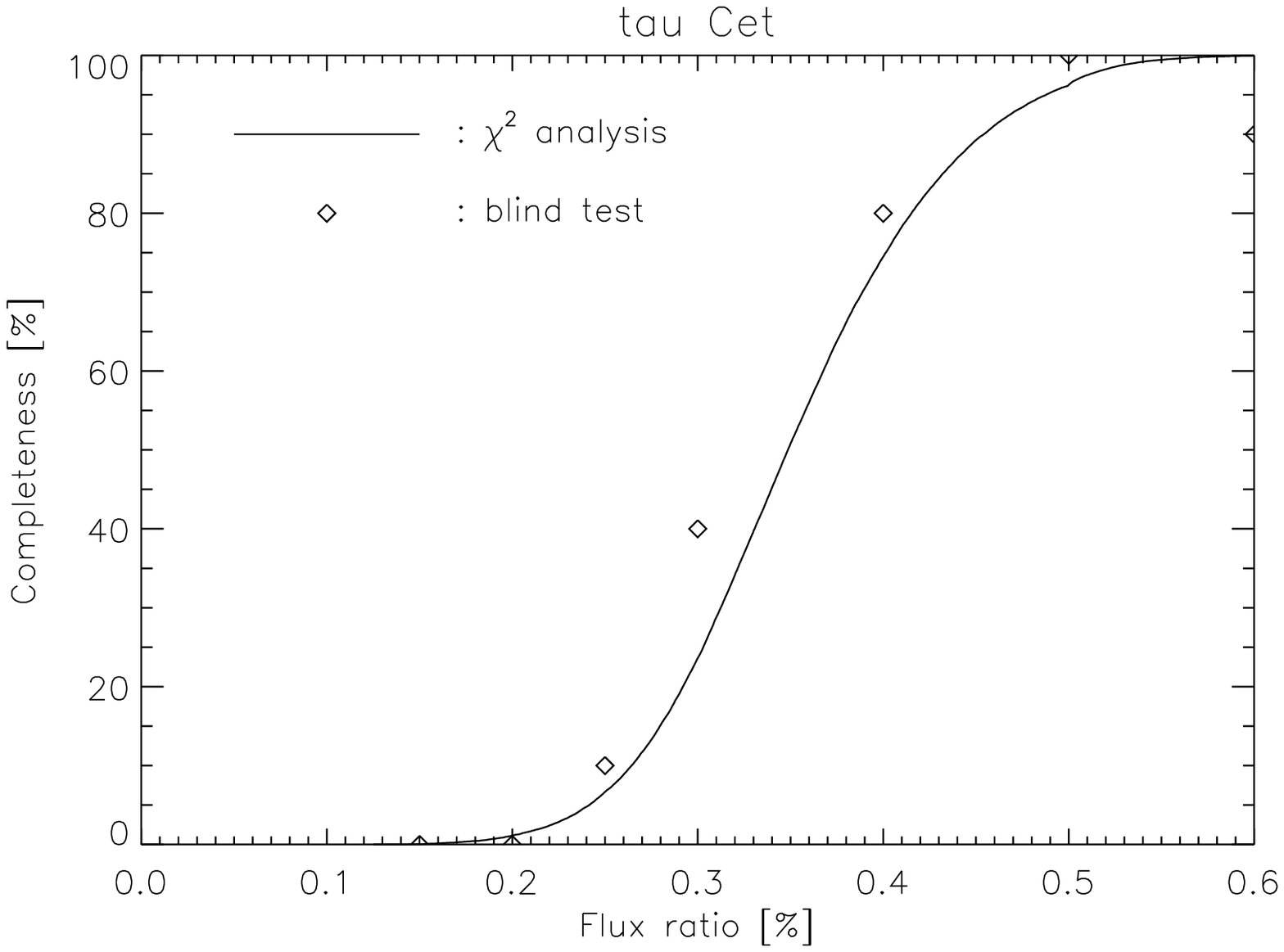} \includegraphics{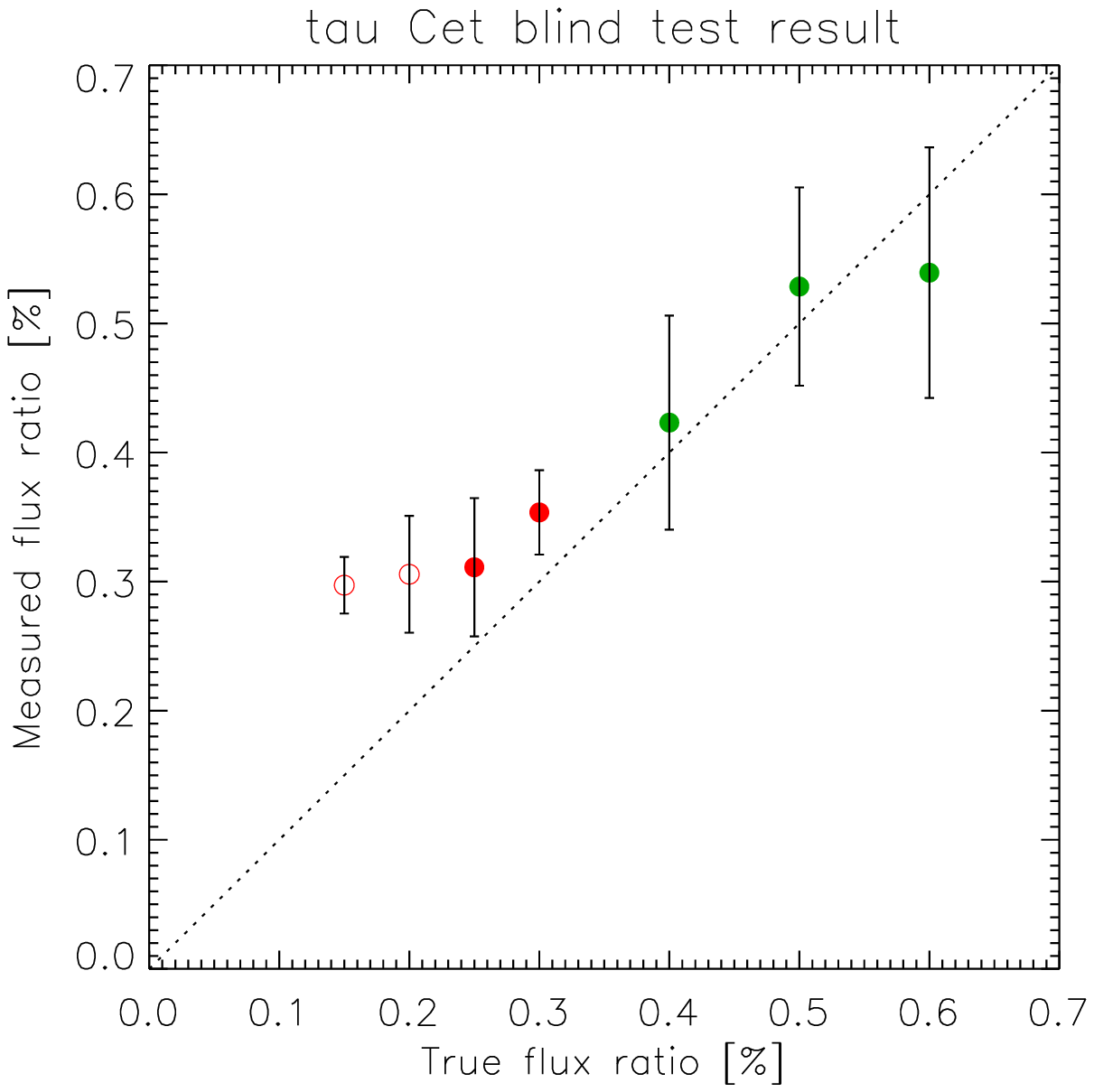}}
\caption{Blind test result for Fomalhaut (top) and tau Cet (bottom). \textit{Left.}  Cumulated histogram of the $3\sigma$ detection limit for companions located within the 100\,mas search region, based on the $\chi^2$ analysis. Over-plotted in diamonds are the results of the double-blind tests discussed in Section~\ref{sub:blind}. \textit{Right.} Measured vs.\ true flux ratio for the blind tests. Filled points are used when 50\% or more of the companions have been found at the good position and with the good flux ratio (empty points otherwise). Green points are used when the mean significance of the best fit is larger than $3\sigma$ (red points otherwise). The error bars represent the dispersion of the best-fit flux ratios.}
\label{fig:blind_test}
\end{figure*}

The blind tests were arranged in series of ten test cases for several values of the flux ratio. A total of 21 different flux ratios ranging between 0.05\% and 0.6\% were investigated in the case of Fomalhaut, while seven different flux ratios between 0.15\% and 0.6\% were used in the case of tau Cet. The results of the blind tests are illustrated in Fig.~\ref{fig:blind_test}. The left-hand side plots show the fraction of companions that was correctly identified (position and flux ratio) as a function of the flux ratio (i.e., the detection ``completeness''). The agreement with the $\chi^2$ analysis is excellent in the case of tau Cet, while the blind test suggests a sensitivity slightly better than expected from the $\chi^2$ analysis in the case of Fomalhaut. In the latter case, the median sensitivity within the 5--50\,mas region would be around 0.19\% instead of the expected 0.23\%. The right-hand side plots display the measured flux ratios of the best-fit binary models as a function of the true flux ratios. These plots demonstrate the absence of bias in the determination of the companion brightness. They also confirm that the flux ratio of false detections generally does not exceed 0.2\% and 0.35\%, respectively, for Fomalhaut and tau Cet. 

			\subsubsection{Short integration on Regulus}

\begin{figure*}[t]
\centering
\resizebox{\hsize}{!}{\includegraphics{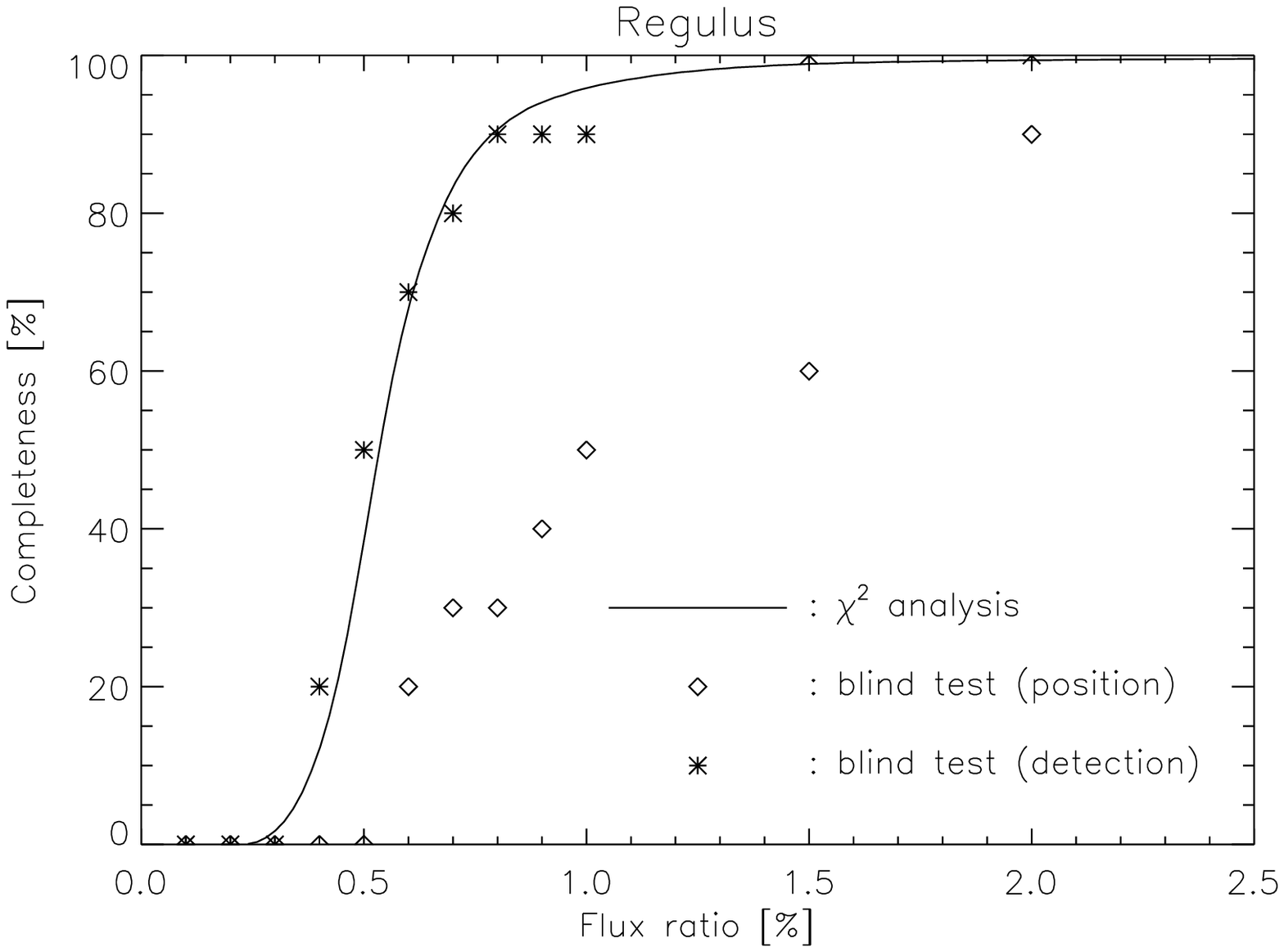} \includegraphics{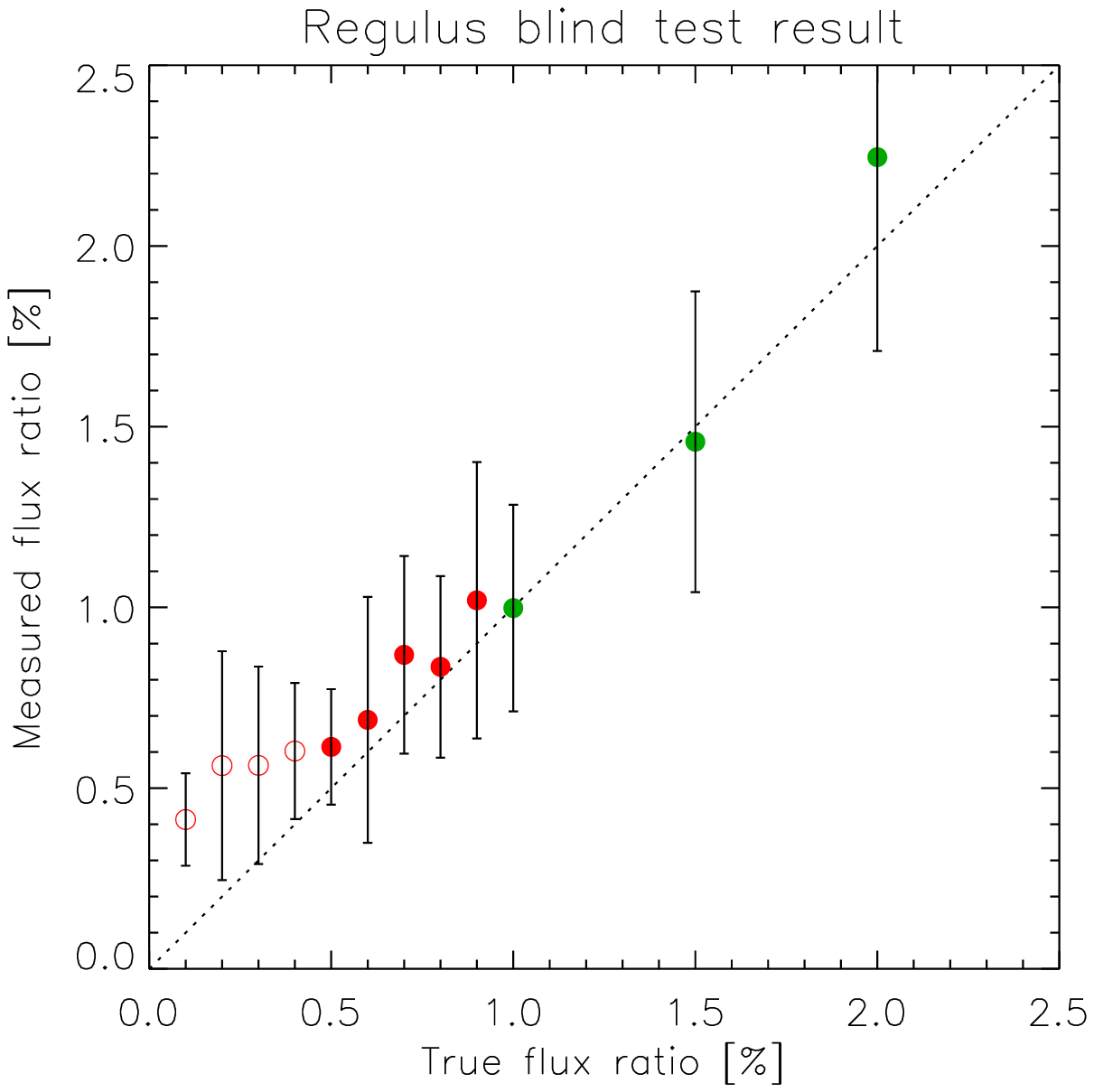}}
\caption{Same as in Fig.~\ref{fig:blind_test} for the case of Regulus. In addition to the immediate blind test results (diamonds), for which the companion is deemed to be detected only when the contrast and position of the global best fit are compatible with the input parameters of the fake companion, we introduce another detection criterion independent of the position of the global best fit (asterisks), which is more suited for short integrations (see text).}
\label{fig:blind_regulus}
\end{figure*}

A similar analysis was performed for Regulus, with contrasts ranging from 0.1\% to 2.0\% for the fake companions. Because the $u,v$ plane coverage is less dense in this case, several replica of the fake companions introduced in the data sets appear frequently in the $\chi^2$ cubes, at various positions in the search region. These can be seen as the side lobes of the instrumental PSF achieved during this snapshot. Due to the statistical noise and systematic errors (which are not fully averaged out due to the low number of OBs/files), the most significant minimum in the $\chi^2$ cube does not always correspond to the actual position of the companion. In Fig.~\ref{fig:blind_regulus}, we have therefore used two definitions of the completeness level: a ``detection'' is reported when a significant minimum (i.e., at more that $3\sigma$) is observed in the $\chi^2$ cube at the companion position, while the ``position'' of the companion is deemed found only when the global minimum in the cube is located less than 1\,mas away from the actual companion position. The left-hand side plot of Fig.~\ref{fig:blind_regulus} illustrates a behaviour that was already noted in the case of del Aqr: even when the detection is (very) clear, the companion position generally remains ambiguous due to the limited $u,v$ plane coverage, and the global best fit is not always found at the right position. This is also the reason for the presence of filled red dots in the right-hand side plot for contrasts ranging from 0.5\% to 0.9\%: the estimated companion position is generally wrong in this contrast range. Nonetheless, the results of the blind test in terms of pure detection (disregarding position) are fully compatible with the $\chi^2$ analysis.

Overall, we have thus used three different approaches to evaluate the sensitivity levels, which all give similar results:
\begin{itemize}
	\item the $\chi^2$ analysis of the original data set, using a $3\sigma$ detection limit,
	\item the detection rate (completeness) in the blind tests,
	\item the typical flux ratio for the non-significant detections ($<3\sigma$) in the blind tests.
\end{itemize}
The agreement between the various approaches further validates the sensitivity levels and confirms that $3\sigma$ is a reasonable significance level for identifying candidate companions. It also validates a posteriori our assumption of Gaussian statistics for our data sets.

    \subsection{Notes on individual targets}

        \paragraph{Fomalhaut}

Fomalhaut is a bright A4V star located at 7.7\,pc. Near-infrared interferometric observations with VLTI/VINCI have recently revealed a K-band excess emission close to the photosphere, with a flux ratio of $8.8\times 10^{-3} \pm 1.2\times 10^{-3}$ \citep{Absil09}. In that paper, it was argued that the excess emission most probably comes from an extended source, rather than from a point-like companion. The absence of closure phase signal in our PIONIER data further confirms this conclusion, and allows the presence of any companion with a flux ratio greater than $2.8\times 10^{-3}$ to be rejected at $3\sigma$ with a 90\% completeness level on a 100\,mas field-of-view. According to the models of \citet{Baraffe98}, this corresponds to a mass limit around $0.175M_{\odot}$ for an age of 290\,Myr \citep{DiFolco04}. We therefore confirm that the previously detected excess emission at K band comes from an extended (and mostly point-symmetric) source.

        \paragraph{tau Cet}

This G8 main sequence star located at 3.6\,pc is also known to have a K-band excess emission, with a flux ratio of $9.8\times 10^{-3} \pm 2.1\times 10^{-3}$ derived from high-precision visibility measurements at the CHARA array with the FLUOR instrument \citep{DiFolco07}. As in the case of Fomalhaut, our PIONIER observations confirm the absence of low-mass companion within the first 100\,mas around tau Cet, with a 90\% completeness level of $4.5\times 10^{-3}$ at $3\sigma$. According to the models of \citet{Baraffe98}, this corresponds to a mass limit around $0.09M_{\odot}$ for an age of 10\,Gyr \citep{DiFolco04}; however, the significance of the best-fit binary model for the tau Cet data set is rather close to $3\sigma$, so that the possibility that a faint companion is actually present cannot be completely ruled out. The best-fit companion would have a flux ratio of $3.0\times10^{-3}$ and would be located at (E: 16.9\,mas, N: 49.5\,mas) from tau Cet. This potential companion would, however, not be bright enough to explain the K-band excess found by \citet{DiFolco07}.

        \paragraph{del Aqr}

This A3 main sequence star, located at $49\pm4$\,pc, is a known astrometric binary with poorly constrained orbital parameters \citep{Goldin07}. \citet{Lagrange09b} shows that its radial velocity is variable and also identifies it as a binary, but fails to constrain the orbit of the companion due to the small time span of their observations ($<400$\,days). With our PIONIER observations, we directly detect the companion for the first time, although with some ambiguity on its position. The orbital parameters cannot be refined based on this sole measurement, and would require the binary to be observed again at several phases along its orbit. However, with this single snapshot, we can readily estimate the spectral type of the companion. The H-band flux ratio amounts to $2.05\times10^{-2} \pm 0.16\times10^{-2}$, which corresponds to $\Delta H=4.22\pm0.09$. Taking the magnitude \citep[$H=3.14\pm0.02$,][]{Bouchet91} and distance of del Aqr into account, the companion has an absolute magnitude $M_H=3.89\pm0.20$. Assuming that the detected point-like source is a main-sequence star orbiting del Aqr, the companion would then have a spectral type around G5V. 

In an attempt to lift the ambiguity on the companion position, we have tried to fit the wavelength-differential squared visibilities together with the closure phases, using the exact same method. Based on our experience with PIONIER data, the differential visibilities are generally less constraining than the closure phases when searching for faint companions, for the reasons explained in Sect.~\ref{sec:analysis}. However, the additional constraints brought by the differential visibilities seem to identify the middle position in Fig.~\ref{fig:probadelaqr} as the true solution. The angular separation and position angle would then be $41.02 \pm 0.37$ mas and $44.2\pm0.5$ degrees, respectively. The ratio of probabilities between the three possible solutions is, however, not high enough to give a definitive answer.

        \paragraph{Regulus}

Our observations rule out the presence of companions with flux ratio greater than $7.9\times 10^{-3}$ at $3\sigma$ with a 90\% completeness level on a 100\,mas field-of-view around this B7 main sequence star. According to the models of \citet{Baraffe98}, this corresponds to a mass limit around $0.62M_{\odot}$ for an age ranging between 50 and 90\,Myr \citep{Che11}.

    \subsection{Limitations and possible improvements}

The error bars on the closure phase estimated by \texttt{pndrs} take into account two main contributions: (i) the statistical dispersion of the data within the 100 scans in each individual file, and (ii) the fluctuations of the transfer function estimated from the calibration measurements. Up to now, the latter has generally been dominant in the error bar budget, so that our observations are not expected to be photon-noise limited. However, based on the PIONIER data obtained so far, we note that the sensitivity to off-axis companions generally improves when increasing the integration time and when observing brighter targets, which are two characteristics of photon-noise limited observations. Determining whether our observations are actually photon-noise limited or not is beyond the scope of this paper.

One crucial aspect for high-precision interferometric observations in general is the calibration. Although we have not noted suspicious variations in the closure phase tranfer function estimated from the calibration stars (as long as the weather conditions were decent), there might be some systematic effects when the calibration stars are located too far from the science target, especially at high air mass. Another possible question is the presence of unknown faint companions around the chosen calibrators. Otherwise, calibration is usually not an issue: the closure phases of all our calibrators are generally consistent with zero within the error bars, and also with the science target's closure phases in case of a non-detection.

Our experience with the PIONIER data shows that four to five OBs of good quality obtained at different hours angles are required to get robust results with our search method. While two or three OBs are generally enough to reach a reasonably good dynamic range, the positions of the detected companions frequently remain ambiguous due to insufficient $u,v$ plane coverage. Furthermore, accumulating several OBs reduces the effect of possible systematics such as drifts in the closure phase at the edges of the wave band (which we could see in some data sets). The typical observing sequence to perform the astrometry of faint companions around bright unresolved stars therefore consists in five observations of the target interspersed with observations of reference stars, using at least two different references to reduce the systematic effects related to calibration. A large-scale survey for companions can, however, rely on two to three calibrated observations, which nowadays takes about one hour with PIONIER.

The ultimate limitation to the dynamic range of PIONIER cannot be predicted yet, because on one hand, the level of the systematic errors cannot be precisely estimated from our limited data set, and because, on the other, the instrument and data reduction software are expected to undergo significant upgrades in the future. In particular, we are currently investigating the possibility of scanning the fringes more rapidly in the destructive read-out mode of the PICNIC camera, in an attempt to better stabilise the fringes during each individual scan.


\section{Conclusion}

In this paper, we have described our method to search for faint companions around bright stars with the PIONIER instrument at VLTI. We have demonstrated that dynamic ranges up to 1:500 can be reached when performing deep integrations (a few hours, including calibration measurements) on mostly unresolved targets, while dynamic ranges around 1:200 can be routinely achieved with three OBs of good quality, which nowadays takes about one hour. We validated our search method by performing a double-blind test study, where fake companions of various flux ratios were introduced in our data sets.

Our observations of Fomalhaut, tau Cet, and Regulus revealed the absence of companion within the 100\,mas search region, with median dynamic ranges at $3\sigma$ respectively of about 1:430, 1:290, and 1:180. At 90\% completeness, the dynamic ranges become 1:360, 1:220, and 1:130. In the first two cases, these non-detections confirm that the near-infrared excess emissions of about 1\% previously detected around these two stars with high-precision visibility measurements at the VLTI and CHARA arrays are related to an extended source of emission rather than a point-like source. For del Aqr, we obtained a direct detection of a faint companion (flux ratio of $2.05\times10^{-2} \pm 0.16\times10^{-2}$) located at about 40\,mas from the primary star, which was already known from astrometric and radial velocity measurements but was poorly constrained up to now.

Based on these results and on the foreseen near-term developments of PIONIER, we expect that dynamic ranges up to 1:1000 should be reachable in the future. Whether PIONIER will eventually be able to meet its most ambitious goal, i.e., reaching the required dynamic range ($\sim$1:5000) to directly detect hot giant exoplanets around unresolved target stars, is not clear yet.

\begin{acknowledgements}
PIONIER is funded by Universit\'e Joseph Fourier (UJF, Grenoble-1) with the programme TUNES-SMING, the Institut de Plan\'etologie et d'Astrophysique de Grenoble (IPAG, ex-LAOG), and the Institut National des Science de l'Univers (INSU) with the programmes ``Programme National de Physique Stellaire'' and ``Programme National de Plan\'etologie''. PIONIER was developed by the CRISTAL instrumental team of IPAG in collaboration with R.~Millan-Gabet (NExScI) and W.~Traub (JPL). The authors want to warmly thank the VLTI team. We acknowledge support from the French National Research Agency (ANR) through project grant ANR10-BLANC0504-01. This work made use of the Smithsonian/NASA Astrophysics Data System (ADS) and of the Centre de Donn\'ees astronomiques de Strasbourg (CDS).
\end{acknowledgements}

\bibliographystyle{aa} 
\bibliography{PIONIER_companions_rev} 

\end{document}